\documentclass[longauth,singlecolumn]{aa}
\usepackage[varg]{txfonts}
\usepackage{graphicx}
\usepackage{epstopdf}
\usepackage{subfig}
\usepackage{color}
\usepackage{array}

\usepackage{natbib}

\begin{document}

\captionsetup[figure]{labelformat=simple, labelsep=period}
\captionsetup[table]{labelformat=simple, labelsep=period}

\newcommand{\msun}{\rm{M}_\odot}
\newcommand{\mste}{\textit{M}_{\star}}
\newcommand{\mstar}{\textit{M}^{*}}
\newcommand{\rtwo}{\textit{r}_{200}}
\newcommand{\rmtwo}{\textit{r}_{-2}}
\newcommand{\br}{\it{\beta (r)}}
\newcommand{\sSFR}{\textit{sSFR}}
\newcommand{\SFR}{\textit{SFR}}
\newcommand{\fc}{\textit{f}_C}
\newcommand{\fM}{\textit{f}_M}
\newcommand{\zphot}{\textit{z}_{phot}}
\newcommand{\zspec}{\textit{z}_{spec}}
\newcommand{\re}{\textit{R}_{e}}
\newcommand{\rc}{\textit{R}_{C}}
\newcommand{\fis}{\it{\Phi}^{*}}

\title{CLASH-VLT: Environment-driven evolution of galaxies in the z=0.209 cluster Abell 209. 
  \thanks{Based in large part on data collected at the ESO VLT
    (prog.ID 186.A-0798), at the NASA HST, and at the NASJ Subaru
    telescope}}

\author{M. Annunziatella\inst{\ref{MAn},\ref{ABi}}
\and A. Mercurio\inst{\ref{AMe}}
\and A. Biviano\inst{\ref{ABi}} 
\and M. Girardi\inst{\ref{MAn},\ref{ABi}} 
\and M. Nonino\inst{\ref{ABi}} 
\and I. Balestra\inst{\ref{ABi}}
\and P. Rosati\inst{\ref{PRo}}
\and G. Bartosch Caminha\inst{\ref{PRo}}
\and M. Brescia\inst{\ref{AMe}}
\and R. Gobat\inst{\ref{RGo}} 
\and C. Grillo\inst{\ref{CGr}}
\and M. Lombardi\inst{\ref{MLo}}
\and B. Sartoris\inst{\ref{MAn},\ref{ABi},\ref{BSa}}
\and G. De Lucia\inst{\ref{ABi}} 
\and R. Demarco\inst{\ref{RDe}}
\and B. Frye\inst{\ref{BFy}} 
\and A. Fritz\inst{\ref{MSc}}
\and J. Moustakas\inst{\ref{JMo}}
\and M. Scodeggio\inst{\ref{MSc}}
\and U. Kuchner\inst{\ref{BZi}}
\and C. Maier\inst{\ref{BZi}}
\and B. Ziegler\inst{\ref{BZi}}
}

\offprints{M. Annunziatella, annunziatella@oats.inaf.it}

\institute{Dipartimento di Fisica, Univ. degli Studi di Trieste, via Tiepolo 11, I-34143 Trieste, Italy\label{MAn} \and
INAF-Osservatorio Astronomico di Trieste, via G. B. Tiepolo 11, 
I-34131, Trieste, Italy\label{ABi} \and
INAF-Osservatorio Astronomico di Capodimonte, Via Moiariello 16 I-80131 Napoli, Italy\label{AMe} \and
Dipartimento di Fisica e Scienze della Terra, Univ. degli Studi di Ferrara, via Saragat 1, I-44122, Ferrara, Italy\label{PRo} \and
Korea Institute for Advanced Study, KIAS, 85 Hoegiro, Dongdaemun-gu Seoul 130-722, Republic of Korea\label{RGo} \and
Dark Cosmology Centre, Niels Bohr Institute, University of Copenhagen,
Juliane Maries Vej 30, 2100 Copenhagen, Denmark\label{CGr} \and 
Dipartimento di Fisica, Universit\`a degli Studi di Milano, via Celoria 16, I-20133 Milan, Italy\label{MLo} \and
INFN, Sezione di Trieste, via Valerio 2, I-34127 Trieste, Italy\label{BSa} \and
Department of Astronomy, Universidad de Concepci\'on, Casilla 160-C, Concepci\'on, Chile \label{RDe} \and
Steward Observatory/Department of Astronomy, University of Arizona, 933 N Cherry Ave, Tucson, AZ, USA\label{BFy} \and
INAF-IASF-Milano, via Bassini 15, 20133 Milano, Italy\label{MSc} \and
Siena College/Department of Astronomy, 515 Loudon Road, Loudonville, NY, USA\label{JMo}  \and
University of Vienna, Department of Astrophysics, T\"urkenschanzstr. 17, 1180 Wien, Austria\label{BZi} 
}

\date{}

\abstract{The analysis of galaxy properties, such as stellar masses, colors, sizes and morphologies, and the relations among them and the environment, in which the galaxies reside, can be used to investigate the physical processes driving galaxy evolution.}
{We conduct a thorough study of the cluster A209 with a new large spectro-photometric dataset to investigate
possible environmental effects on galaxy properties that can inform us
on galaxy evolution in cluster hostile environments.}
{We use the dataset obtained as part of the CLASH-VLT spectroscopic survey, supplemented with Subaru/SuprimeCam high-quality imaging in BVRIz bands, which yields to 1916 cluster members (50 \% of them spectroscopically confirmed) down to a stellar mass $\mathrm{\mste \, =\,  10^{8.6}\, \msun}$. We determine the stellar mass function
of these galaxies in different regions of the cluster, by separating the
sample into star-forming and passive cluster members. We then
determine the intra-cluster light and its properties.  We also
derive the orbits of low- ($\mathrm{\mste\, \leq \, 10^{10.0}\, \msun}$) and high-mass ($\mathrm{\mste\, > \, 10^{10.0}\, \msun}$) passive galaxies and study the effect of the environment
on the mass-size relation of early type galaxies, selected according
to their Sérsic index, separately for galaxies in the two mass
ranges. Finally, we compare the cluster stellar mass density profile with the number- and total-mass density profiles.}
{The stellar mass function of the star-forming cluster galaxies does not depend on
the environment. The slope found for passive galaxies becomes flatter in the most dense cluster region, which implies that the low-mass component starts to dominate when moving away from the cluster center. The color distribution of the intra-cluster light is consistent with the color of passive cluster members. The analysis of the dynamical orbits of passive galaxies shows that low-mass galaxies have tangential orbits, avoiding small pericenters around the BCG. The mass-size relation of low-mass passive early-type galaxies is flatter than that of high mass galaxies, and its slope is consistent with the slope of the relation of field star-forming galaxies. Low-mass galaxies are also more compact  within the scale radius of $\mathrm{0.65\, Mpc}$. The ratio between the stellar and number density profiles shows a mass segregation effect in the cluster center. The comparative analysis of the stellar and total density profiles indicates that this effect is due to dynamical friction.
}
{
Our results are consistent with a scenario in which the
"environmental quenching" of low-mass galaxies is due to mechanisms
such as harassment out to $\rtwo$, starvation, and ram-pressure stripping 
at smaller radii. This scenario is supported by the
analysis of the mass function, of the dynamical orbits and of the
mass-size relation of passive early-type galaxies in different
cluster regions. Moreover, our analyses support the idea that the
intra-cluster light is formed through the tidal disruption of subgiant
($\mste \, \mathrm{\sim\, 10^{9.5-10.0}\, \msun }$) galaxies. In fact, our results suggest that low-mass galaxies are destroyed by tidal interactions, and that those which avoid small pericenters around the BCG, are influenced by tidal interactions that reduce their sizes. 
We suggest dynamical friction as the process responsible of the observed mass segregation.
}


\titlerunning{A209 stellar mass function}
\authorrunning{M. Annunziatella et al.}

\maketitle

\section{Introduction}
\label{s:intro}

Galaxies in clusters experience many different processes, which
affect their properties and evolution. Among them the environmental
(i.e. nurture) processes play a key role in the evolution of galaxies
\citep{boselli14} especially at the low-mass
end. These mechanisms can be divided into two main classes:
gravitational interactions, including all sorts of tidal interactions,
i.e. galaxy-galaxy mergers, dynamical friction and harassment, and
hydrodynamic interactions, including interactions between the galaxy’s
interstellar medium (ISM) and the hot intergalactic medium, i.e. ram pressure (for a review
see \citealt{biviano2008}). These processes are effective in different
regions of the cluster (e.g. \citealt{treu03}, \citealt{boselli06}). Their main effect is to
halt the star formation of a galaxy, changing it from star-forming (SF) to quiescent. 
Other properties that could be influenced by these
processes are: stellar mass, size, and morphological type. 
The study of the distribution of these properties and the correlation existing among them and with their environment gives information on the relative importance of the processes that affect their evolution. 

One possible way to understand the influence of these processes
on the evolution of galaxies in clusters is to search for effects
of local environment on the distribution of galaxy stellar
masses ($\mste$), i.e. the stellar mass function (SMF), and also
compare the SMF of galaxies in clusters to that in the field.
To disentangle the effects of the environment on the SMF from the effects related to the morphology-density relation, i.e. the fact that the fraction of different morphological types changes with local density \citep{dressler1980, postman2005}, one should study the SMF separately for different types of galaxies. 
So far, the determination of the SMF has been mainly based on data in the field \citep{ilbert2013}, while
the evolution with redshift (\textit{z}) and the local environment of the SMF of different galaxy types in clusters is instead still poorly studied.
On the other hand, the study of the luminosity function (LF) is better  known. 
By separating the contribution of early- and late-type galaxies using color indices, \cite{popesso06} show that the r-band LF of local late-type galaxies can be fitted with a single Schechter function, while the LF of early-type galaxies requires a double Schechter function. They also show that the faint-end slope of the LF of early-type galaxies exhibits a significant and continuous variation with the environment, flattening in the central region of the cluster. 
They interpret these results as an effect of galaxy transformation from star forming to quiescent systems through harassment in the periphery and dwarf tidal disruption in the core.\\
\cite{vulcani2012, vulcani2013} have found that global environment does not affect the SMF of different galaxy types. Furthermore, they did not detect any difference between the SMFs of the global population of galaxies within and outside the virial radius of galaxy clusters. \\
However, it is also important to stress that the effects of the environment could be evident only below a given mass limit for a given redshift.
For example, at $z\, \mathrm{\sim 1}$, \cite{vanderburg2013} do find a difference in the shape of the SMF of clusters and of the field below $\mathrm{10^{10} \, \msun}$. \\
In \cite{annunziatella2014}, we have studied the effect of environment on the SMF of SF and passive galaxies in the galaxy cluster MACS 1206.2-0847 (hereafter M1206) at the mean redshift of $z\, \mathrm{=\, 0.44}$, by use of a large fraction ($\mathrm{\sim}$ 1/3) of spectroscopically confirmed cluster members, down to $\mste \,\mathrm{=\, 10^{9.5} \, \msun}$.
We have found that the SMF of both galaxy types does not change across the virial radius\footnote{The virial radius $\rtwo$ is the
  radius of a sphere with mass overdensity $\Delta$ times the critical
  density at the cluster redshift. Unless otherwise stated, we adopt
  $\Delta=200$ throughout this paper.} of that cluster. 
However, we did detect a difference on the SMF of passive galaxies in the innermost and most dense region of the cluster in the form of a lack of galaxies with $\mste \,\mathrm{<\, 10^{10.5} \, \msun}$. Our hypothesis is that these galaxies in the  very center of the cluster have been destroyed by tidal interactions with the cluster potential and have enriched the intra-cluster light. \\

Another way to constrain the effect of different environmental processes on the evolution of galaxies is to analyze the distribution of galaxies sizes and the relation with the stellar mass. It is known that different types of galaxies, namely early types (ETGs) and late types show different dependencies between size and stellar mass \citep{shen2003}. In particular, the mass-size relation of ETGs is steeper than the one found for late type galaxies. How the environment may change the size distribution of the galaxies, instead, is still matter of debate. In the local Universe, \cite{poggianti2010} claim that, at fixed stellar mass, cluster ETGs are slightly smaller than field ones. On the other hand, \cite{huertas-company2013} did not detect any relation between galaxy sizes and environment. Many of the physical processes that can affect the properties of cluster galaxies are likely to be related to their orbits. Galaxies on
radial orbits, characterized by small pericenters, can experience
denser environments than galaxies on more tangential orbits. This can lead to their more rapid destruction \citep{taylor2004}. It has
been suggested by \cite{solanes2001} that among Virgo cluster galaxies, those with evidence of gas stripping are on more radial orbits than the others.

\begin{table}[ht]
\centering
\begin{tabular}{l>{\bfseries}c>{\bfseries}c>{\bfseries}lc}
 \hline
BCG $(\alpha,\delta)_{\mathrm{J2000}}$ & $01^{\mathrm{h}}31^{\mathrm{m}}52\fs54, -13\degr 36\arcmin 40\farcs4$ \\
Mean redshift & $0.2090 \pm 0.0004$\\
Velocity dispersion [km~s$^{-1}$] & $1320_{-67}^{+64}$ \\
Virial radius $\rtwo$
 [Mpc] & $2.13 \pm 0.05$ \\
\hline
\end{tabular}
\caption{Main properties of the cluster Abell 209. The radius $\rtwo$ corresponds to the radius of 
a sphere whose mass overdensity is 200 times the critical density at the cluster redshift. The value of the redshift is taken from \cite{mercurio2003_2}. The velocity dispersion is calculated on our dataset (Sartoris et al. in prep.), while the virial radius is taken from \cite{merten2014}. }
\label{t:props}

\end{table}

In this paper, we investigate the SMF of passive and SF galaxies in the cluster Abell\,209 (hereafter, A209) at $z\,\mathrm{=\,0.209}$, as a function of the local environment and down to a mass limit of $\mathrm{10^{8.6}\, \msun}$. We also analyze the distribution of the sizes of passive early-type galaxies in different mass ranges and in different cluster regions. Finally we determine the orbits for cluster passive galaxies in two
mass bins.

These analyses are based on an unprecedented large fraction ($>$ 50\%) of $\mathrm{\sim  1000}$ spectroscopically confirmed cluster members collected at the ESO-VLT, as a part of the ESO Large Programme 186.A-0798 (hereafter CLASH-VLT, P.I. Piero Rosati, \citealp{ros14}). This is a spectroscopic follow-up of a subset of 14 southern clusters from the CLASH (``Cluster Lensing And Supernova survey with Hubble'') (see \citealt{postman2012}).

The main properties of A209 are listed in Table~\ref{t:props}. We perform the analysis of the SMF for two reasons: to check whether the results found for the dynamically relaxed cluster M1206 holds also for a cluster with evidence of substructures \citep{mercurio2003} and, to study possible redshift evolution effects comparing the results obtained in the two clusters, which are separated by $\sim$2 Gyr in cosmic time. For A209, we also study the intra-cluster light (ICL) component in order to give some hints on its progenitors and to understand how it is related to the evolution of the galaxies in the cluster. Some of the environmental dependent mechanism that shape the SMF of cluster member galaxies, in fact, can also be responsible for the formation of this diffuse component. The ICL was first discovered by \citet{zwicky1951} in the Coma cluster. 
It origins from tidal interactions among the galaxies or with the cluster potential during the cluster assembly. It is still a matter of debate 
which galaxies contribute more to the ICL. From a theoretical point of view one of the proposed leading scenarios is that this component comes from tidal disruption of intermediate mass galaxies  ($\mathrm{10^{10\, - \, 11} \, \msun}$, \citealt{ contini+14}). This scenario has also been supported by some observational works (\citealt{annunziatella2014, Presotto+14, demaio2015, montes2014}). However, some authors (e.g. \citealt{conroy2007}) suggest that the central ICL, associated to the BCG, could form from destruction of dwarf galaxies. The joint analysis of the ICL and the SMF of cluster members can shed some light on this topic.\\
The paper is organized as follows. In Sect.~\ref{s:data}, we describe the data sample and explain how we assign the membership to the galaxies in the field of view. In Sect. ~\ref{s:mass} we derive the stellar masses of galaxies cluster members.
 In  Sect.~\ref{s:smf}, we estimate the SMF for cluster members and examine its dependence on galaxy type and environment. In Sect.~\ref{s:icl}, we give a brief description of the method used to derive the map of the intra-cluster light and study some properties of the ICL in A209. In Sect. ~\ref{s:orbits}, we perform the analysis of the orbits of passive galaxies. In Sect.~\ref{s:sms} we analyze the mass-size relation of passive  galaxies in two regions of the cluster.
 In Sect.~\ref{s:smfrac}, we show the stellar mass density and number density profiles compared with the total mass density profile obtained from an independent lensing analysis \citep{merten2014}. In Sect.~\ref{s:disc} and 
 ~\ref{s:conc}, we discuss our results and draw our conclusions. \\
Throughout this paper, we use $\mathrm{H_0 \, = \, 70\, km\, s^{-1}\, Mpc^{-1} }$,
$\mathrm{\Omega_M \, = \, 0.3, }$ and $\mathrm{\Omega_{\Lambda} \, =
  \, 0.7 }$. At the cluster redshift the scale is 206 kpc $\mathrm{arcmin^{-1}}$. All the magnitudes used in this work are referred to the AB system.

\section{The data sample}
\label{s:data}

A total of 11 LR-Blue masks were acquired with VLT/VIMOS, as part of
the CLASH-VLT ESO large programme, using four
separate pointings, each with a different quadrant centered on the
cluster core. The LR-Blue masks cover the spectral range 370-670 nm
with a resolving power R = 180, for a total exposure time of 10.5h. Spectra are classified according to the reliability of the redshift
(see \citealt{biviano2013}) in: “secure” (QF = 3; 1946 objects) 
with a probability of $>$99.99\%, “based on emission-lines” (QF = 9;
110 objects) with a probability of $\sim$ 92\%, and
“likely” (QF = 2; 492 objects) with $\sim$ 75\% probability.

Membership for objects with spectroscopic redshift is assigned according to the “peak+gap” (P+G) method \citep{fadda96}, already applied to M1206 (\citealt{biviano2013}, \citealt{girardi15}). This method identifies a peak in the redshift distribution using a 1D adaptive kernel algorithm, then rejects galaxies that in overlapping radial bins have line of sight velocities very different from the mean velocity of the cluster. 

The final spectroscopic dataset consists of 2548 reliable redshifts, observed as part of CLASH-VLT, in addition to 110 taken from \cite{mercurio2003_2} and \cite{mercurio2008}. In total, 1116 are confirmed members.

\begin{figure}
\includegraphics[width=\columnwidth]{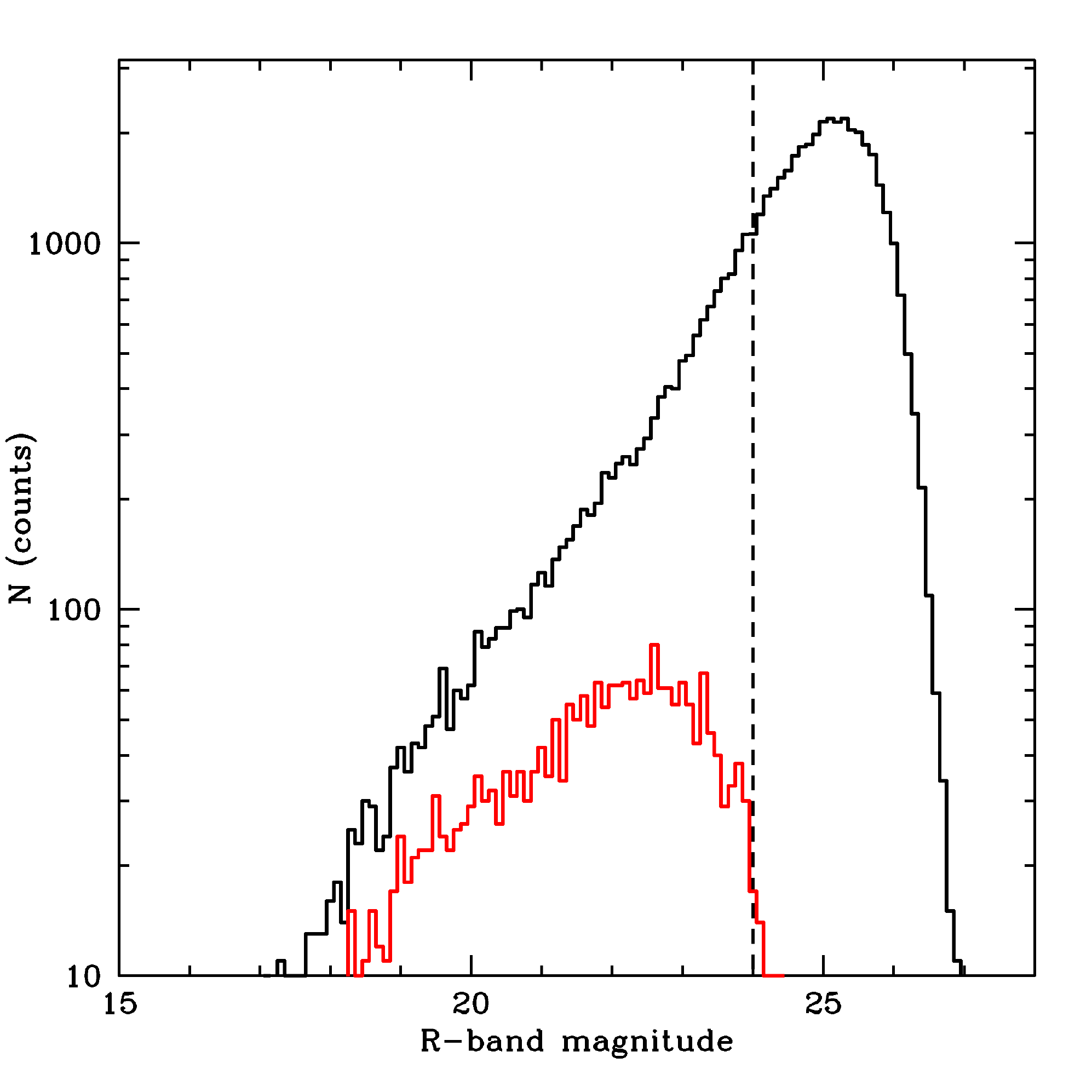}
\caption{Histogram of the $R_C$-band magnitude of the 51166 objects in the photometric (black line) sample and of the 2622 sources with reliable redshifts (red line).}
\label{mhist}
\end{figure}

Photometric data were obtained with the SUBARU/Suprime-Cam from
SMOKA (\citealp{baba2002}) covering a 30$^\prime$ $\times$ 30$^\prime$ field of view, with
total exposure times of 2400\,s, 1800\,s, 2400\,s, 1320\,s and 4800\,s
in the B, V, R$_{\mathrm{c}}$, I$_{\mathrm{p}}$ and z band, respectively. The typical seeing in the final sky-subtracted images varies from 0.486 arcsec in the R band up to 0.77 in the B band with a pixel scale of 0.2 arcsec. Details on the data reduction
can be found in \cite{nonino2009}. The photometric catalogues were extracted using the software SExtractor (\citealp{ber96}) in conjunction with PSFEx (\citealp{ber11}, \citealp{ann13}).
By using a method based
on neural networks (\citealp{brescia2013}), we obtain the photometric redshift, $\zphot$ for all galaxies for which we have the magnitudes in all the observed bands. In this paper, we consider only galaxies down to the limit of the observed spectroscopic sample: $\rc \, =\, 24.0$ mag (see Fig. ~\ref{mhist}).

\begin{figure}[ht]
 \centering
\includegraphics[width=\columnwidth]{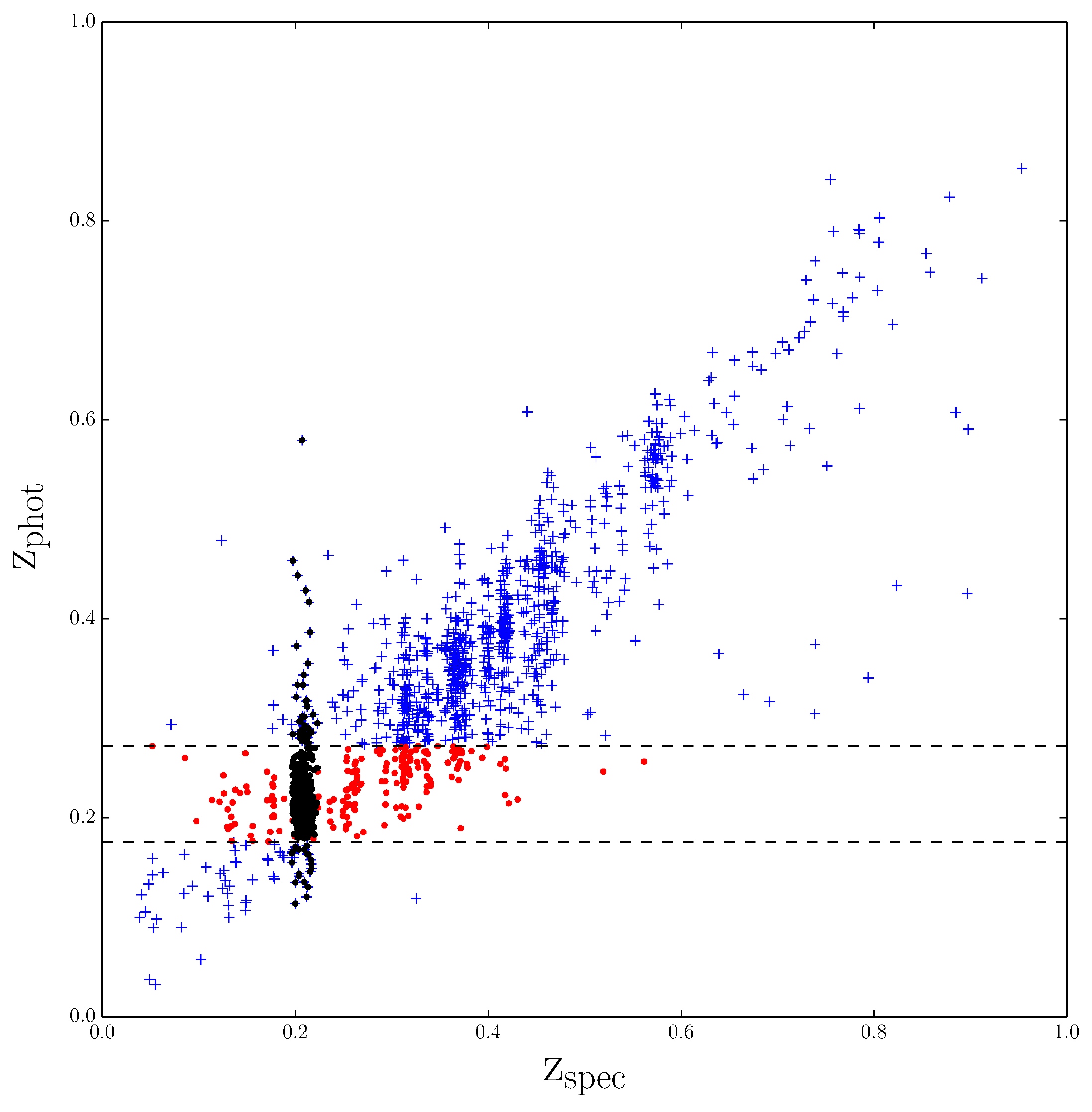}
\caption{Photometric vs. spectroscopic redshifts for galaxies in the
  cluster field and in the magnitude range $\mathrm{18\, \leq \, }\rc
   \mathrm{\, \leq \, 24}$ mag. 
    Black dots are spectroscopically confirmed members.
    The two dashed lines represent the chosen criteria in photometric redshifts to classify the cluster members. 
    Galaxies within this range of photometric redshift are colored in red. 
    Galaxies with $\zphot$ in this range and without spectroscopic information are classified as members.
    Blue crosses are galaxies outside both the spectroscopic and photometric selections. }
\label{f:zzp}
\end{figure}

To determine if galaxies with $\zphot$ can be classified as members, we investigate the diagram shown in Fig. \ref{f:zzp}.
We use the objects with spectroscopic redshift to find the best selection criteria to identify cluster member galaxies. Then, we apply a cut in the photometric redshift. 
We select as cluster members those galaxies with $\mathrm{0.175 \,< \,}\zphot \mathrm{\,<\, 0.272}$ (see horizontal lines in Fig.~\ref{f:zzp} ).
With these cuts in $\zphot$ we classify correctly 92\% of the spectroscopic members, minimizing the number of interlopers.
The sample of photometrically selected members contains 1805 galaxies down to the magnitude limit of $\rc \mathrm{\, =\, 24.0}$ mag.
The selection of members based on photometric redshifts is not as reliable as that based on spectroscopic redshifts. We take this  into account as explained in Sect.~\ref{ss:compl}.

\subsection{Completeness and membership corrections}
\label{ss:compl}
\begin{figure}[hŧ]
 \centering
\includegraphics[width=1.\columnwidth]{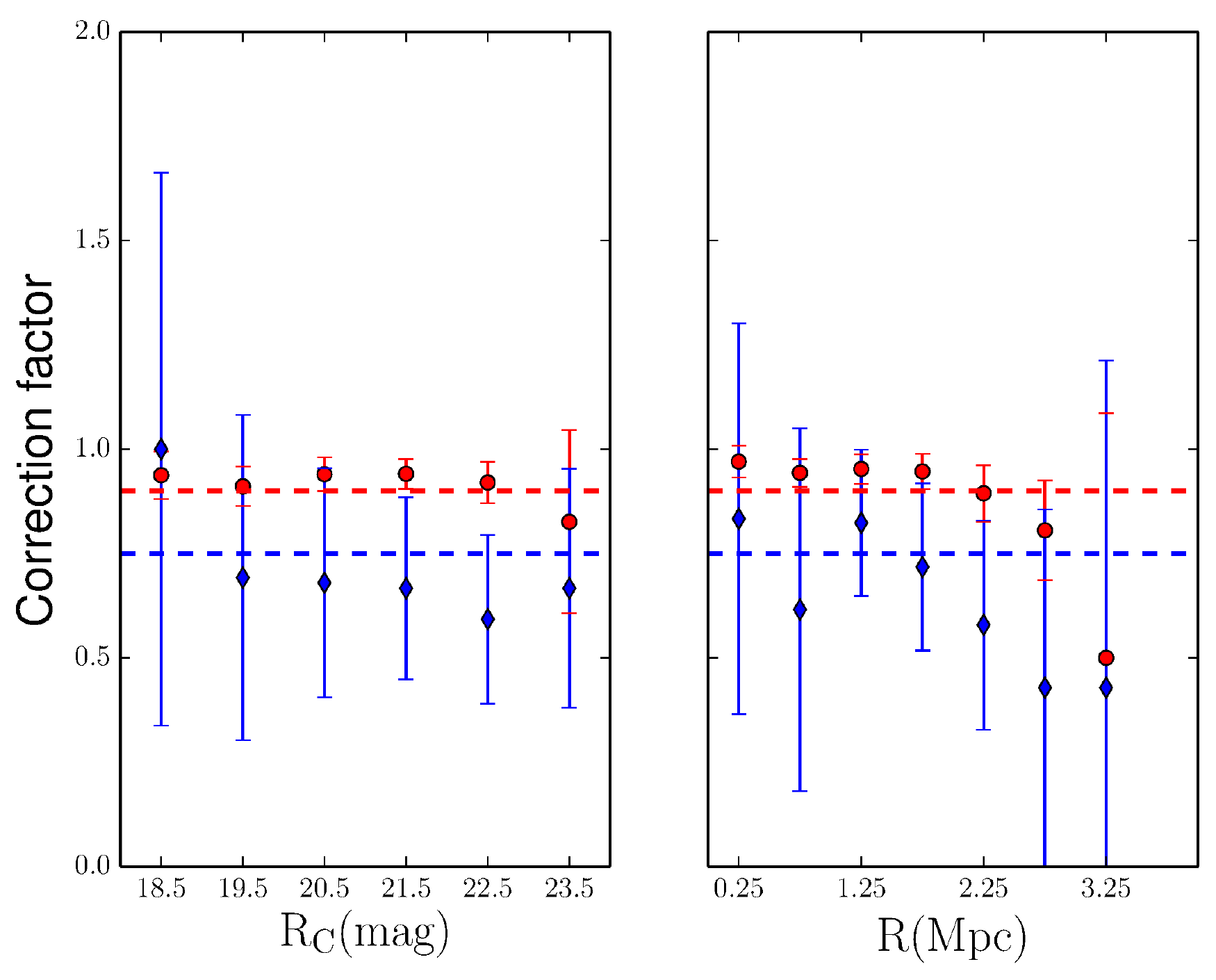}
\caption{Correction factor ($\mathrm{\fM}$) for membership selection as a function of $R_C$ magnitude (left panel) and clustercentric distance (right panel), for red (dots) and blue (diamonds) galaxies.}
\label{f:fm}
\end{figure}

Our sample needs to be corrected for two different factors. 
One is the incompleteness of the sample with photometric redshift with respect to the photometric sample, since to obtain $\zphot$, we require that the galaxies have to be detected in all the five SUBARU bands.
The other takes into account possible discrepancies between $\zphot$ and $\zspec$ in the photometric selection of members, and we call it membership correction. 
In order to account for these corrections, 
we apply the same method as in \cite{annunziatella2014}. \\
The photometric completeness,  \textit{C}, is defined as the ratio between the number of galaxies with photometric redshift and the total number of galaxies in the $\rc$ band.
The completeness,   \textit{C}, has been evaluated for the whole sample and separately for red and blue galaxies, where we use the threshold value $B\mathrm{\, - \, }R_C \mathrm{\, =\, 1.5}$ (calculated on the basis of aperture magnitudes within 5\arcsec) to discriminate between the two samples. This value for the color is chosen to match the limit in specific star formation rate (\sSFR) used to distinguish between Passive and SF galaxies (see Sect. \ref{s:mass}).
 \textit{C} is approximately constant, $\mathrm{\geq \, 80 \%}$, down to $R_C \mathrm{\, \leq \, 24\, mag}$ independently from magnitude or galaxy type. Therefore, we adopt a constant mean value  \textit{C}= 0.81 and a correction factor, $\fc \mathrm{\, =} $ 1/ \textit{C} = 1.23. \\
To obtain the membership correction factor, we evaluate the completeness and the purity of the sample of galaxies classified as members based on their photometric redshift. We define the completeness as the ratio between the number of galaxies identified as members both spectroscopically and photometrically and the number of spectroscopic members $\mathit{C_M} \mathrm{\, =\, } \mathit{N_{pm\, \cap \, zm}}/\mathit{N_{zm}}$. 
The purity is defined as the ratio between the number of photometric members that are confirmed also spectroscopically and the number of photometric members that have $\zspec$, $\mathit{P} \mathrm{\, = \,} \mathit{N_{pm\,\cap
    \, zm}}/\mathit{N_{pm\, \cap \, z}}$. The membership correction factor is defined as $\fM \mathrm{\, \equiv\, } \mathit{P/C_M}$.
In Fig. \ref{f:fm} we analyze the dependence of $\fM$ from galaxy type, magnitude and distance from the cluster center. We find that the correction factor depends on galaxy type, while it is independent from both magnitude and clustercentric distance. We use two different values for red and blue galaxies in the entire magnitude and radial range,  $\fM$ = 0.9 and 0.75, respectively.\\
The final correction factor we applied is given by the product $\fc\,  \cdot  \, \fM$ ($\sim$ 1.125 and 0.937 for passive and SF galaxies, respectively).

\section{Measurements of stellar mass}
\label{s:mass}
\begin{figure}
 \includegraphics[width=\columnwidth]{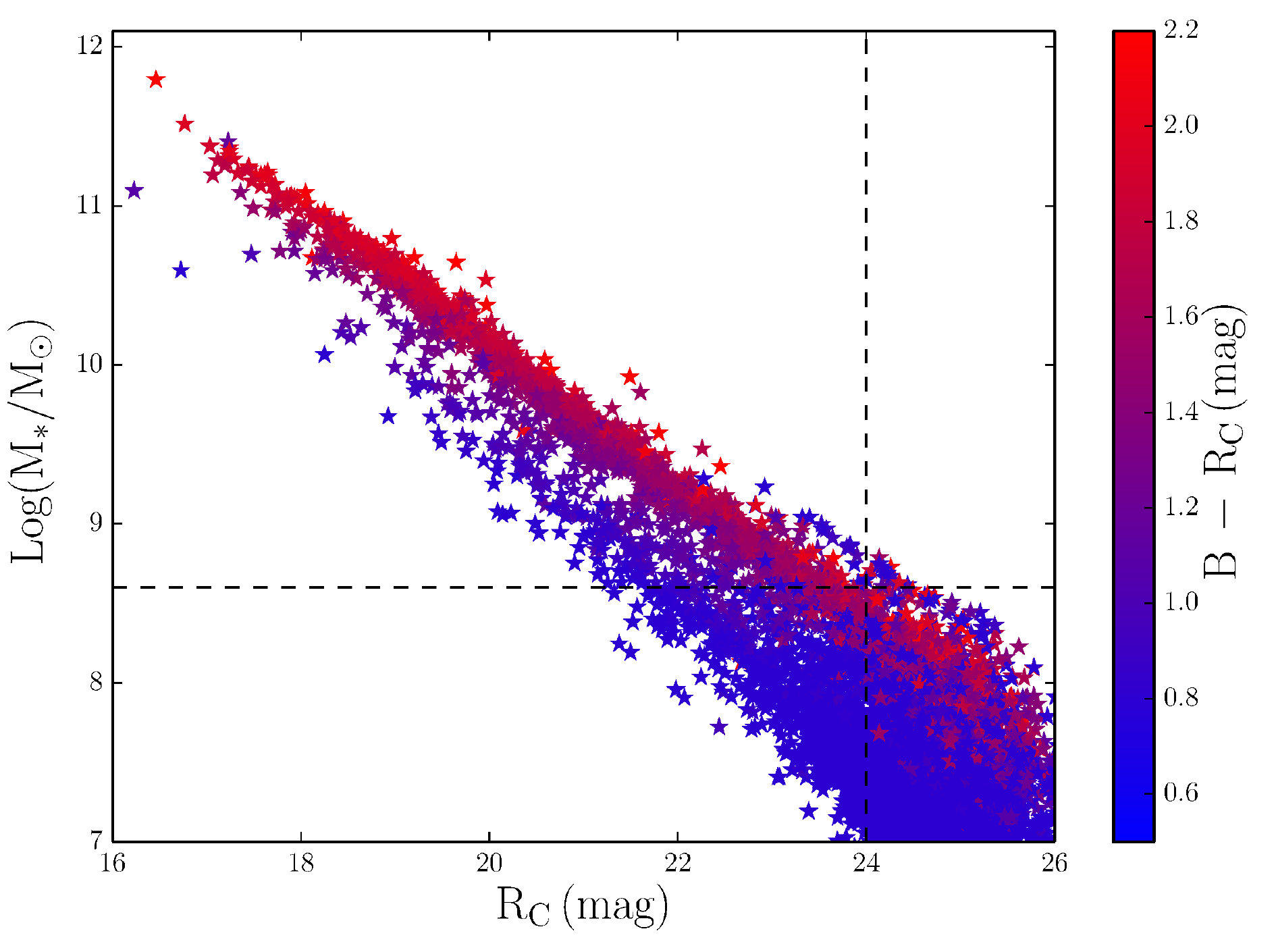}
 \caption{Relation between the stellar mass and the $\rc$ magnitude. The points are color coded according to the $B\mathrm{\, -\, }\rc$ color. Dashed vertical and horizontal lines are the magnitude and mass limits, respectively.}
  \label{f:Mrc}
  \end{figure}

We use the spectral energy distribution (SED) fitting method to derive the stellar masses of the cluster members galaxies. We make use of the code \texttt{MAGPHYS} (\citealt{dacunha2008}). For each galaxy, \texttt{MAGPHYS} finds the template, at the redshift of the galaxy, that best reproduces the observed galaxy fluxes. The templates are assembled 
from stellar population synthesis models (\citealt{bruzual&charlot2003} or Bruzual \& Charlot 2007) with a \cite{chabrier2003}  stellar initial mass function and a metallicity value in the range 0.02--2 $\mathrm{Z_{\odot}}$. We adopt the \cite{bruzual&charlot2003} library of models. The two libraries differ in the treatment of the thermally pulsating asymptotic giant
branch stellar phase which is not relevant at the mean redshift of the cluster A209 (\citealt{zibetti2013}). In order to obtain the SED, \texttt{MAGPHYS} parametrizes the star formation history of a galaxy with a continuum, $\mathit{SFR} \mathrm{\propto e^{-\mathit{\gamma} t}}$, and superimposed random bursts. The star formation time-scale $\mathit{\gamma}$ is distributed according to
the probability density function $\mathrm{p(\mathit{\gamma}) =1\, -\,
  \tanh(8\mathit{\gamma} \, - \, 6)}$, which is uniform between 0 and 0.6
$\mathrm{Gyr^{-1}}$ and drops exponentially to zero at 1
$\mathrm{Gyr^{-1}}$. The age of the galaxy is free to vary with an upper limit imposed by the age of the Universe at the considered redshift. \\
In order to find the best-fit model, \texttt{MAGPHYS} uses a Bayesian approach. As outputs, it produces the probability distribution function of all the parameters and their best-fit values. For the parameters we are interested in we adopt the median value as the more robust estimate with a $\mathrm{1\sigma}$ interval given by the difference between the 16\% and 84\% percentiles of the probability distribution. The mean value for the 1$\sigma$ uncertainty on the stellar mass is $\sim 0.07$ dex. \\
In \cite{annunziatella2014}, we checked that the value of the stellar mass is independent from the specific algorithm of SED fitting (provided one considers the same stellar IMF). However, we found that using only optical data  \texttt{MAGPHYS} tends to overestimate the mass of a galaxy. This can be attributed to the fact that the public \texttt{MAGPHYS} libraries contain an excess of star-forming dusty models, with respect to passive ones (E. da Cunha, priv. comm.). This means that if the fluxes of a galaxy can be fitted by a dusty model or by a truly passive one, the median value is always biased towards the dusty model. In order to avoid this systematic error we make use of a non public library of templates heavily biased to fit old stellar populations with very little star formation (provided by E. da Cunha). 
We proceed as follows. We perform the SED fitting for all the galaxies in our sample using the standard libraries. Among its output parameters, \texttt{MAGPHYS} provides also an estimate of the specific star formation rate (i.e., star formation rate per unit mass, \sSFR $\equiv$ \SFR/$\mste$). We examine the values obtained for the \sSFR~ and check that this parameter has a bimodal distribution, with a local minimum at \sSFR $\,= \, 10^{-10}$ yr$^{-1}$ (see also \citealt[][and references therein]{laralopez2010}). We then classify as passive galaxies those with values of \sSFR $\,< \, 10^{-10}$ yr$^{-1}$. SF galaxies, instead, have \sSFR $\, \geq \, 10^{-10}$ yr$^{-1}$
At this point we fit again the passive galaxies using the non public library of models.
The limiting magnitude of $\rc\mathrm{\, =\, 24}$ mag corresponds roughly to a completeness
limit of $\mathrm{10^{8.6} \, \msun}$ and to a surface brightness of 24.7 mag/$\mathrm{arcsec^2}$, for passive galaxies, according to the mass vs $\rc$ dispersion of Fig.~\ref{f:Mrc} and to the measured galaxy sizes (see Sect~\ref{s:sms}). 
The mass limit for SF galaxies would be lower. However, we choose the same value for passive galaxies in order to be consistent when comparing the SMF of different galaxy type. 
With this mass limit, our final sample consists of 1916 galaxies, of which more than half are spectroscopically confirmed members.

\section{The stellar mass function}
\label{s:smf}

The SMF of cluster members is obtained by considering the number of galaxies in each stellar mass bin weighted for the completeness and the membership correction factors. The errors on the counts are estimated with the bootstrap procedure (Efron and Tibshirani 1986).
The SMF can be analytically described by a \citet{schechter1976} function
\begin{equation}
\Phi (\log M) = \ln(10) \, \fis \,
\left(\frac{M}{\mstar}\right)^{1+\alpha}
\exp\left(-\frac{M}{\mstar}\right) \, d(\log M),
\label{eq:schec}
\end{equation} 
where $\alpha$ is the slope of the low-mass end,  $\mstar$ is the exponential cutoff, and $\Phi^*$ is the normalization. We fit the Schechter function to the $\mste$ distribution down the mass limit $\mathrm{10^{8.6}\, M_{\odot}}$ using the maximum likelihood technique \citep{MK86}. The use of this method implies that only two parameters are left free in these fits, while the normalization is fixed by imposing that the integral of the SMF in the chosen mass range is equal to the total number of galaxies of the considered sample corrected for the weight factor. In these fits each galaxy has a weight equal to the product $\mathit{\fc \cdot \fM}$ defined in Sect.~\ref{ss:compl}. 
We have checked that the contribution of the errors on the stellar mass value is smaller than the statistical errors associated to each of the parameters of the SM. The  correction factor in Fig. \ref{f:fm}, as shown, is independent from magnitude or radial distance. Therefore, the errors on this factor 
would influence only the normalization of the SMF. The only two cases in which we take into account the normalizations are when studying the total SMF (Sect. ~\ref{ss:type}) and when estimating the contribution of galaxies of different mass to the ICL (Sect. ~\ref{ss:icl_p}). To estimate the contribution of the errors in these two cases, we repeat our calculations using $\fM\mathrm{\, \pm\, }\Delta \fM$. The results we obtain are statistically consistent with those obtained using $\fM$. Thus, for our analysis we do not consider the errors on the membership correction factor.
The cumulative SMF for all galaxies in A209 is given in  Fig. \ref{f:smf} (violet line). \\
We are interested in studying how the SMF depends on galaxy type or environment in order to shed  light on the different processes that can affect the evolution of galaxies in clusters. 
We want to compare only the shape of the SMF of different samples, so we renormalize the SMFs by dividing them for the total number of objects in each sample (except in Fig.~\ref{f:smf}). In order to establish if two SMFs are statistically different we compare the shape parameters ($\alpha \mathrm{\, and \,} \mstar$) without taking into account the normalization values.
To highlight the difference between two SMFs, we also perform the Kolmogorov-Smirnov (hereafter, K-S) test (e.g., \citealt{press2013}). This is a non-parametrical test which determines if two samples are drawn from the same parent population by evaluating the maximum distance between their two cumulative distributions. We use a modified version of the K-S test in order to include the weights on stellar mass \citep{annunziatella2014}.

\begin{table*}[ht]
\centering
\caption{Best-fit Schechter function parameters}
\label{t:sbf1}
\begin{tabular}{lcccccc}
 \hline
Galaxy type & $\alpha$ & $\mathrm{log(\mstar/\msun)}$ \\
\hline
Passive  & -1.09 $\mathrm{\pm}$ 0.02 & 11.10 $\mathrm{\pm}$ 0.05 \\
SF  & -1.63 $\mathrm{\pm}$ 0.03 & 11.02 $\mathrm{\pm}$ 0.30 \\
All  & -1.17 $\mathrm{\pm}$ 0.02 & 11.15 $\mathrm{\pm}$ 0.05 \\
\hline
\end{tabular}
\end{table*}

\subsection{Different galaxy types}
\label{ss:type}

\begin{figure}[ht]
\centering \includegraphics[width=\columnwidth]{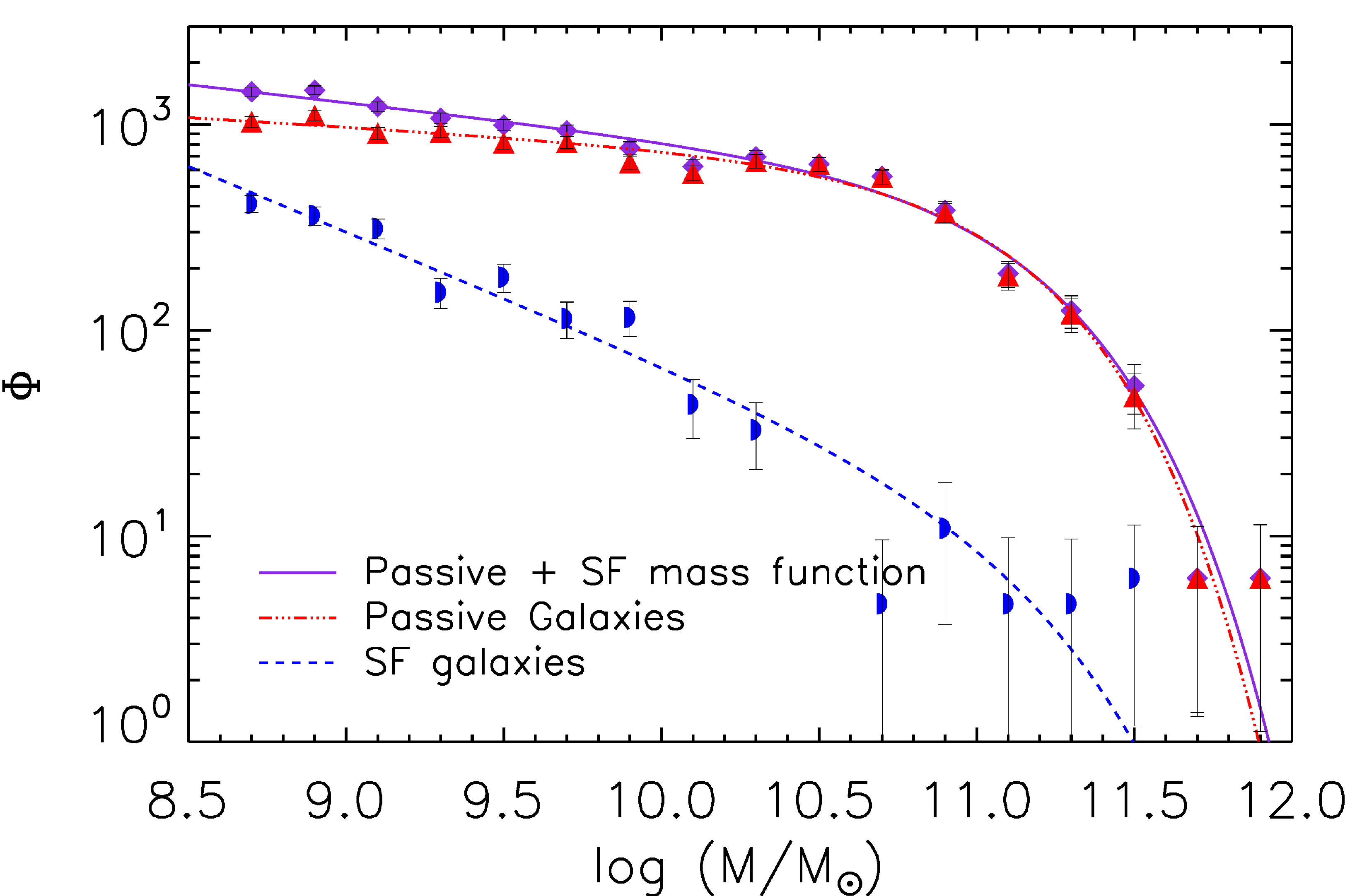}
  \caption{SMFs of SF and passive cluster member galaxies. The best-fit Schechter functions are shown with the blue dashed line for SF galaxies and red triple-dot-dashed line for passive galaxies, while the number counts (divided for the bin size) are the blue demi-circle and red triangle, respectively.
   Violet diamonds are the counts obtained for all cluster members and the violet solid line represents the sum of the two SMFs. Counts are evaluated in bins of 0.2 dex in $\mste$. The 1~$\sigma$ errors on the counts have been
       estimated via the bootstrap resampling procedure.  }
  \label{f:smf}
  \end{figure}

\begin{figure}[ht]
 \includegraphics[width=\columnwidth]{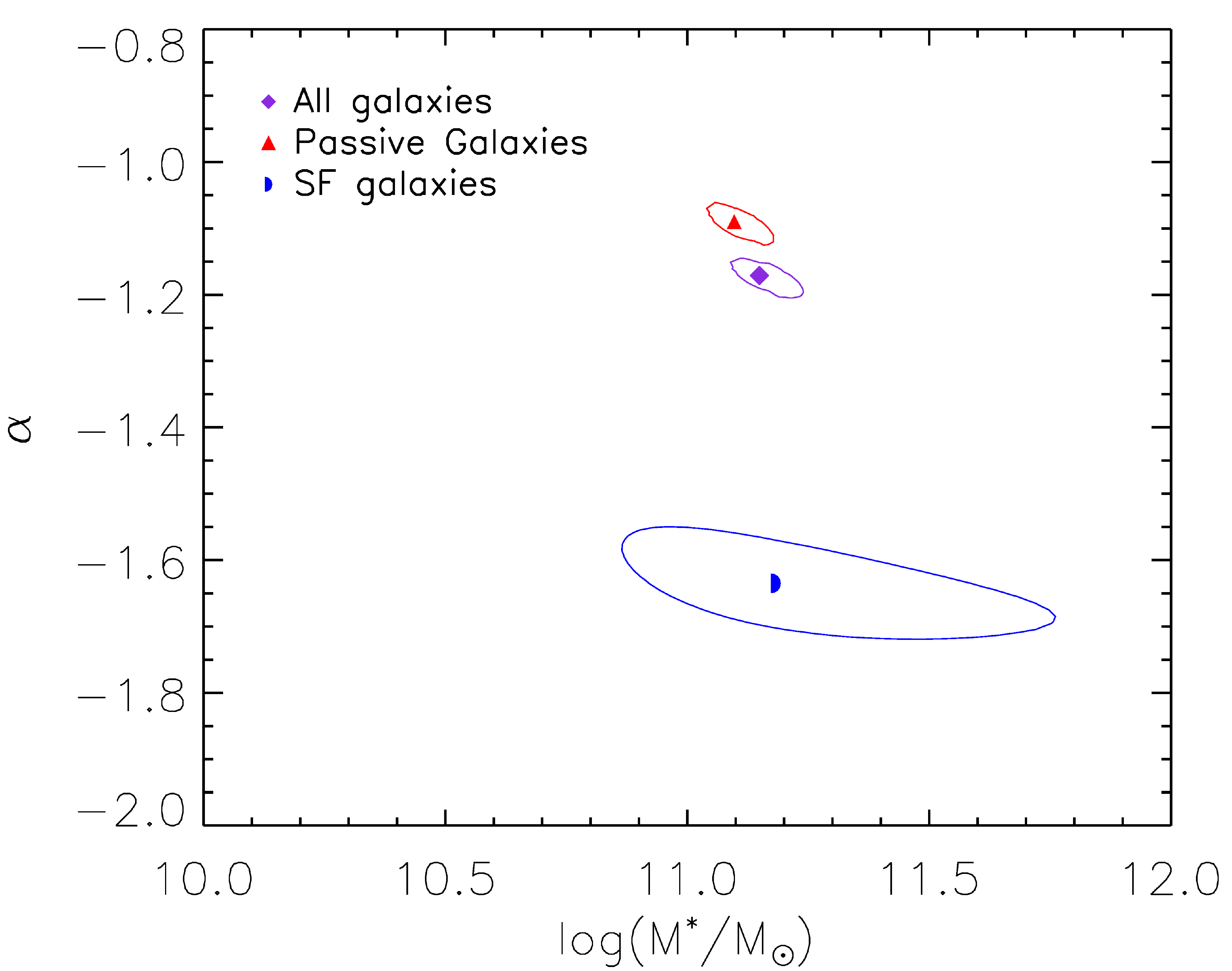}
 \caption{Best-fit values of the Schechter parameters $\alpha$ and $\mstar$ with their 1~$\sigma$ likelihood contours.}
  \label{f:sbf}
  \end{figure}

The SMFs of separately SF and passive cluster member galaxies are shown in Fig. \ref{f:smf}. The best-fit values of $\alpha \mathrm{\, and \,} \mstar$ with their 
1$\sigma$ errors are listed in Table \ref{t:sbf1} and shown in Fig. \ref{f:sbf}. The uncertainties are obtained  by marginalizing over the normalization parameter. The SMFs of, separately, SF and passive cluster members are adequately fitted by a single Schechter function. In order to assess the goodness of the fits, we compare the mass distribution of galaxies in the considered sample with the best-fit Schechter function using the K-S test. We perform this test for all the SMFs considered in this paper, except for the SMF of all cluster members, always obtaining P-values $\mathrm{> \, 10\%}$. 
Therefore, we can not reject the null hypothesis that the two distributions are statistically identical.
The SMF of passive and SF cluster members is given by the sum of the SMFs of separately SF and passive galaxies. The dependence of the SMF from galaxy type is statistically significant. This is evident from the comparison of the shape of the two SMFs but also from the results of the K-S test (see Table~\ref{t:ks}). The difference between the mass distribution of passive and SF galaxies holds not only when considering the whole cluster, but also in different cluster regions. 
From Fig. \ref{f:smf} we see that the passive SMF dominates over the SF one at all masses down to the completeness limit in A209. In \cite{annunziatella2014}, we found that the two SMFs of SF and passive galaxies for the cluster M1206 at $z\mathrm{ \sim \, 0.44}$ crossed at the value $M_{cross}\, \mathrm{\sim 10^{10.1}\, \msun}$, if considering the whole cluster region, and $M_{cross}\, \mathrm{\sim 10^{9.5}\, \msun}$, if we restricted the analysis inside the virial region of the cluster. The difference between what happens in the two clusters can be partially explained by the Butcher–Oemler effect (\citealt{butcher1978}), according to which galaxy clusters at intermediate redshift contain a larger fraction of blue galaxies than lower redshift ones. On the other hand, there is a striking difference between the slope of the SMF of passive galaxies that we find in this work and the one we found for passive galaxies in M1206 ($\mathit{\alpha_{passive\, M1206} \, =\,- 0.38 \, \pm \, 0.06}$). We check that this  difference in slope is not due to the different completeness limit that we reach in the two works. The difference between the SMF of passive galaxies in the two clusters at different redshifts is qualitatively in agreement with the results shown in \citet{rudnick2009}. In their paper, the authors found that the luminosity functions of passive galaxies showed a strong evolution with redshift in terms of an increase of faint galaxies relative to bright ones towards lower redshifts. 
Another feature of the passive galaxy SMF is the dip visible in the galaxy counts at $\mathrm{\sim \, 10^{10.10}\, \msun}$. The presence of this dip was also found in the luminosity function of red-sequence galaxies of this cluster (\citealt{mercurio2003}) at the magnitude in the R band $\mathrm{\sim \, 19.8\, mag}$ which corresponds (according to the fit in Eq. ~\ref{e:rcMfit}) to our mass value. This dip is probably an indication that there are two populations of passive galaxies, one of lower masses probably originated from the quenching of SF galaxies, and another of higher masses. To verify this we fit a double Schecter to our data in Sect. ~\ref{ss:ds}.

\begin{table}[ht]
\centering
\caption{Best-fit Schechter function parameters for different environments}
\label{t:sbf}
\begin{tabular}{llcc}
 \hline
Galaxy type & Environment & $\alpha$ & $\mathrm{log(\mstar/\msun)}$\\
\hline
\\
SF & $\mathrm{R > \rtwo}$ & -1.58 $\mathrm{\pm}$ 0.08 & 10.95 $\mathrm{\pm}$ 0.30\\
SF & $\mathrm{R \leq \rtwo}$ & -1.69 $\mathrm{\pm}$ 0.07 & 11.42 $\mathrm{\pm}$ 0.45\\
\\
Passive & $\mathrm{R > \rtwo}$ & -1.09 $\mathrm{\pm}$ 0.04 & 11.09 $\mathrm{\pm}$ 0.09\\ 
Passive & $\mathrm{R \leq \rtwo}$ & -1.09 $\mathrm{\pm}$ 0.02 & 11.10 $\mathrm{\pm}$ 0.05\\
\\
Passive & Region (a) & -1.01 $\mathrm{\pm}$ 0.06 & 11.21 $\mathrm{\pm}$
0.12\\
Passive & Region (b) & -1.09 $\mathrm{\pm}$ 0.04 & 11.09 $\mathrm{\pm}$
0.08\\
Passive & Region (c) & -1.10 $\mathrm{\pm}$ 0.03 & 11.09 $\mathrm{\pm}$
0.08\\
Passive & Region (d) & -1.10 $\mathrm{\pm}$ 0.03 & 11.03 $\mathrm{\pm}$ 0.07\\
\hline
\end{tabular}
\end{table}

\begin{table}[ht]
\centering
\caption{Results of the K-S tests}
\label{t:ks}
\begin{tabular}{lcr}
 \hline
Compared samples & N1, N2 & Prob. (\%) \\
 \hline
\\
\multicolumn{3}{c}{Type dependence}\\
\\
Passive vs. SF in the cluster  & 1580, 336 & $<0.01$ \\
Passive vs. SF in Region (b) & 425, \phantom{0}34 & $<0.01$\\
Passive vs. SF in Region (c) & 488, \phantom{0}75 & $<0.01$\\
Passive vs. SF in Region (d) &  \phantom{0}75, 219 & $<0.01$\\
\\
\multicolumn{3}{c}{Environment dependence - SF galaxies}\\
\\
SF within and outside $\rtwo$ & 161, 175 & $>10$\\
SF in Regions (b) and (c) & \phantom{0}34, \phantom{0}75  & $>10$\\
SF in Regions (c) and (d) & \phantom{0}75, 219 & $>10$\\
\\
\multicolumn{3}{c}{Environment dependence - passive galaxies}\\
\\
Passive within and outside $\rtwo$ & 438, 408 & $>10$\\
Passive in Regions (a) and (b) & 160, 425 & 8\\
Passive in Regions (b) and (c) & 425, 488 & $>10$\\
Passive in Regions (c) and (d) & 488, \phantom{0}57 & $>10$\\
\\
\hline
\end{tabular}
\tablefoot{The values N1 and N2 are the number of galaxies in each of the compared subsamples. The probabilities in the last columns are referred to the null hypothesis that the stellar mass distributions of the two samples are drawn from the same parent population. A value $\mathrm{\leq}\, 1$\% means that the two distributions are statistically different. Samples with less that 10 objects are excluded from the test.}
\end{table}

\subsection{Effects of global and local environments}
\label{ss:env}

\begin{figure}[ht]
 \includegraphics[width=\columnwidth]{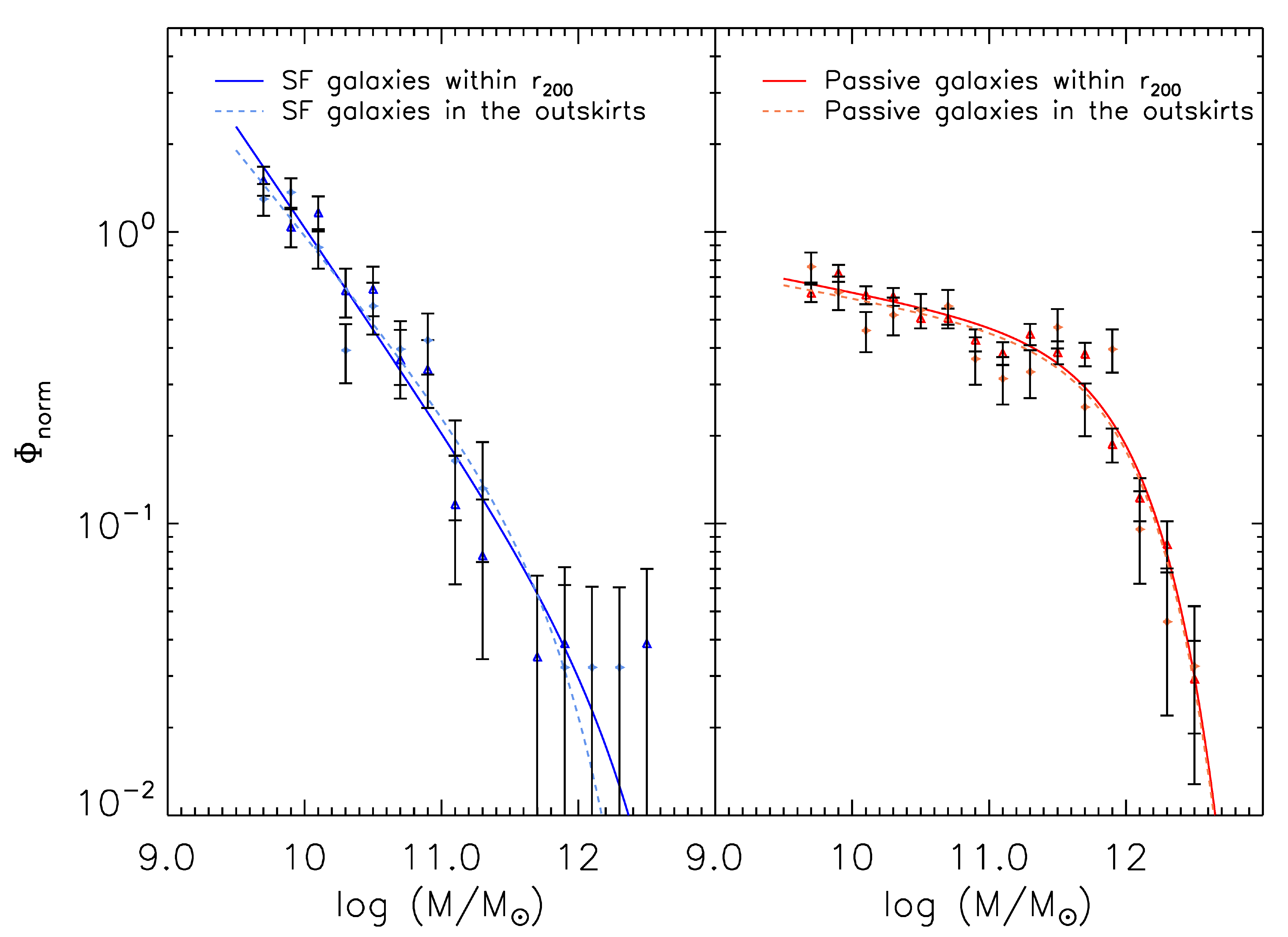}
\caption{SMFs of SF (left) and passive (right) galaxies inside and outside the virial radius of the cluster. The SMFs are normalized to the the total number of galaxies in each sample.}
  \label{f:smf_in_out}
  \end{figure}

\begin{figure}[ht]
\includegraphics[width=\columnwidth]{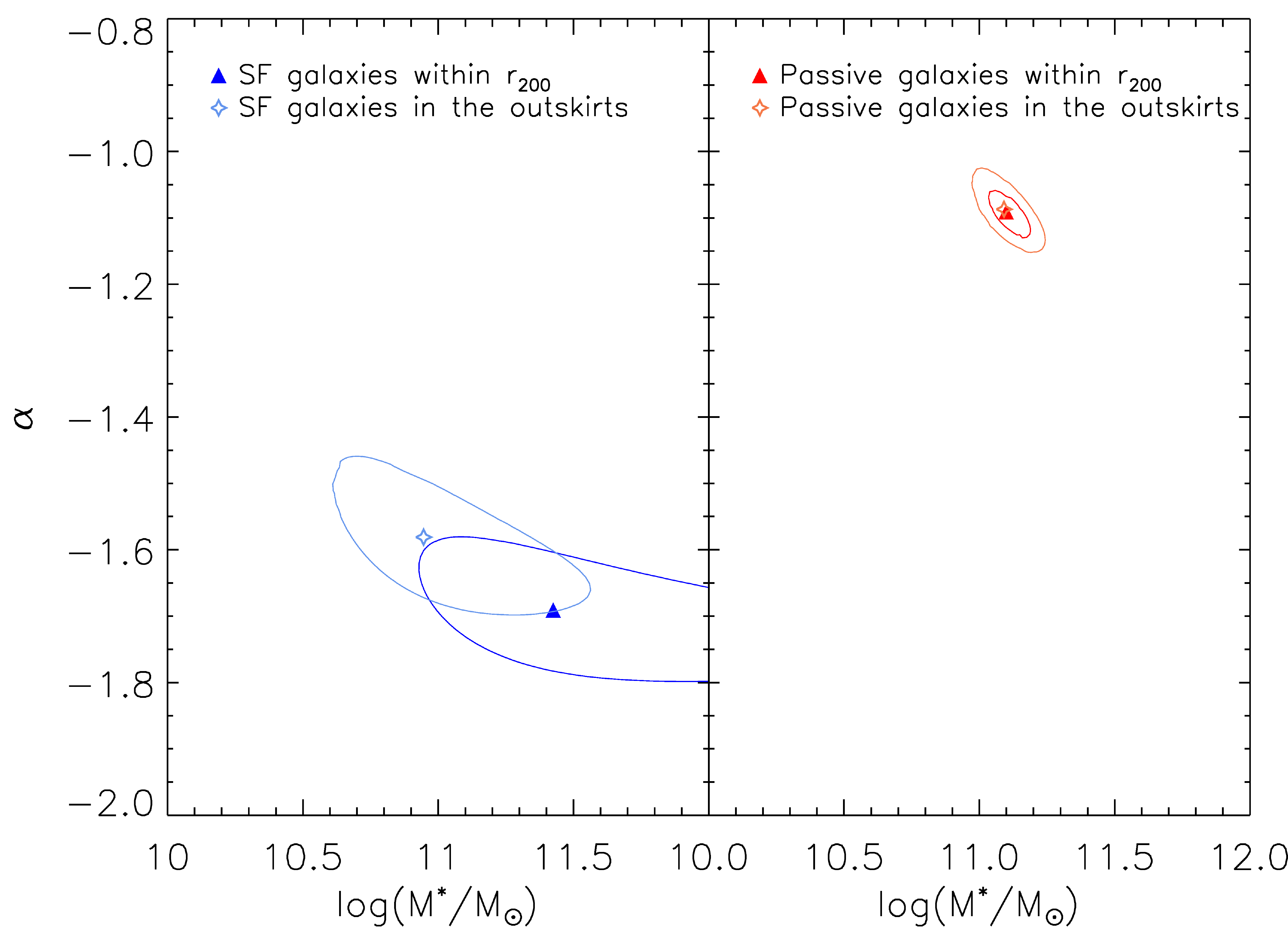}
  \caption{As in  Fig.~\ref{f:sbf}: best-fit Schechter parameters $\mstar$ and
      $\alpha$ with their 1~$\sigma$ contour in different regions of the cluster}
  \label{f:c_in_out}
  \end{figure}

To investigate the environmental dependences of the SMF, we consider separately SF and passive galaxies. 
In this way we can exclude the effects of the dependence of galaxy population from environment (\citealt{dressler1980}). \\
First we consider the SMFs of SF and passive galaxies inside the virial radius of the cluster ($\rtwo$) and outside ($\mathrm{\sim 1.7 \, }\rtwo$).
The SMFs and their best-fit Schechter parameters are shown in Fig.~\ref{f:smf_in_out} and Fig.~\ref{f:c_in_out}.
The SMFs of SF galaxies inside and outside the virial radius are statistically indistinguishable (see the results of the K-S test in Table~\ref{t:ks}). The same result holds for passive galaxies, too.\\

\begin{figure}[ht]
\centering
 \includegraphics[width=\columnwidth]{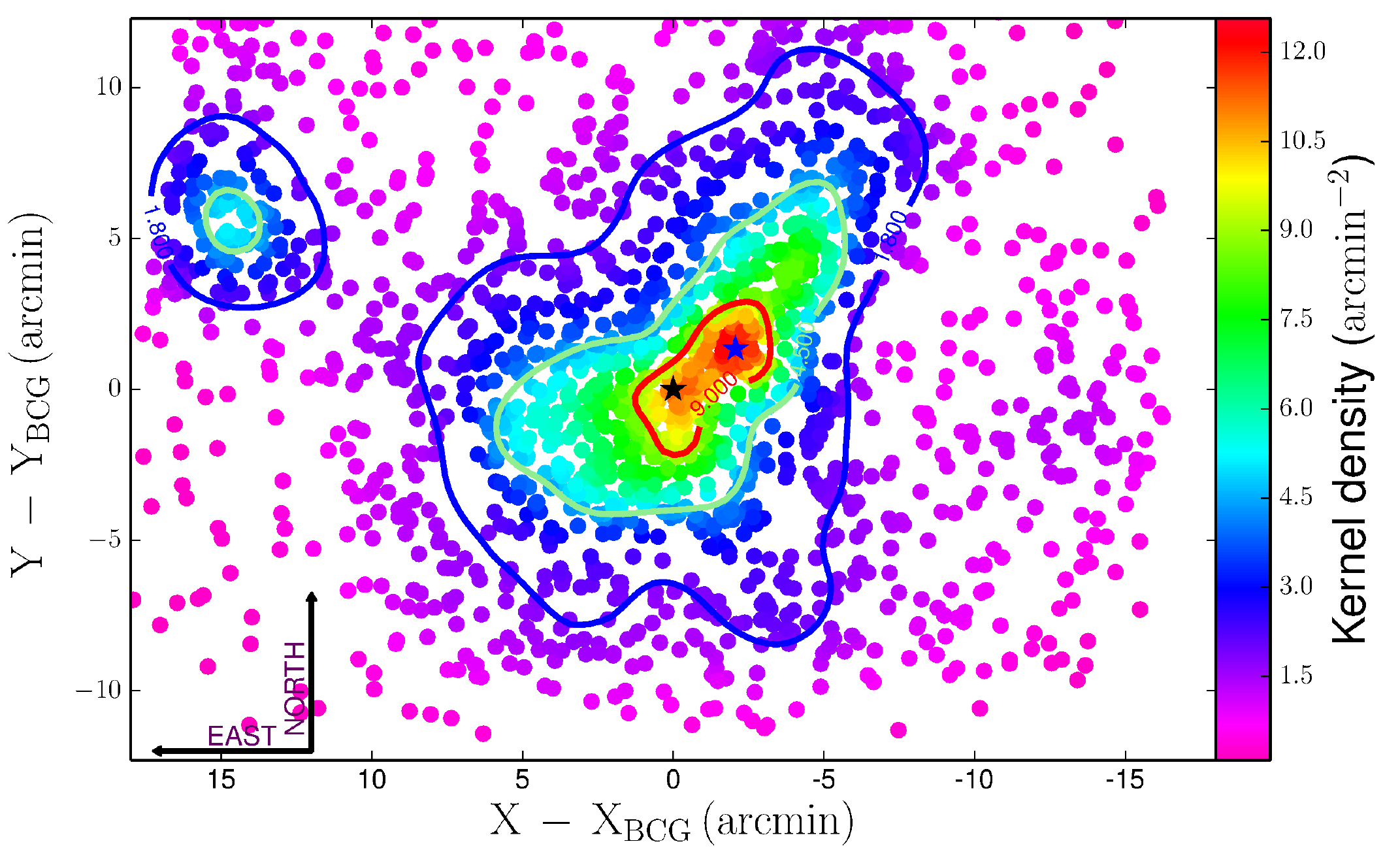}
  \caption{Spatial distribution of cluster members down to the stellar mass limit. The points are colored according to the local number density. Coordinates are referred to the position of the BCG (black star). Red, light-green and blue lines identify galaxies belonging to Regions (a), (b) and (c) defined in the text. Points outside the contours belong to Region (d). The blue star represents the position of the highest density peak.}
  \label{f:dens}
  \end{figure}

We define the `environment' in units of the local density (number of galaxies per $\mathrm{arcmin^{-2}}$).
We decided to use this definition of environment instead of the radial distance from the BCG (\citealt{annunziatella2014}) since A209 is an elongated, dynamically not fully relaxed cluster (Fig. ~\ref{f:dens}) and the regions identified by local density are elongated with respect to those identified by clustercentric radius.

The local density is obtained with a kernel density function, where the projected distribution of the galaxies is smoothed with a Gaussian kernel in an iterative way.
Fig. \ref{f:dens} shows the distribution of galaxies in the cluster. The points are color coded according to the value of the local density ($\mathrm{\Sigma}$). 
We define four different regions according to their number density (in terms of galaxies per arcmin$^{-2}$):
  
\begin{enumerate}[(a)]
 \item $\mathrm{\Sigma >\, 9.0}$.
 \item $\mathrm{4.5 <\, \Sigma\, \le\, 9.0}$.
 \item $\mathrm{1.8 <\, \Sigma\, \le\, 4.5}$.
 \item $\mathrm{\Sigma\, \le\, 1.8}$.
\end{enumerate}

\begin{figure}[ht]
 \includegraphics[width=\columnwidth]{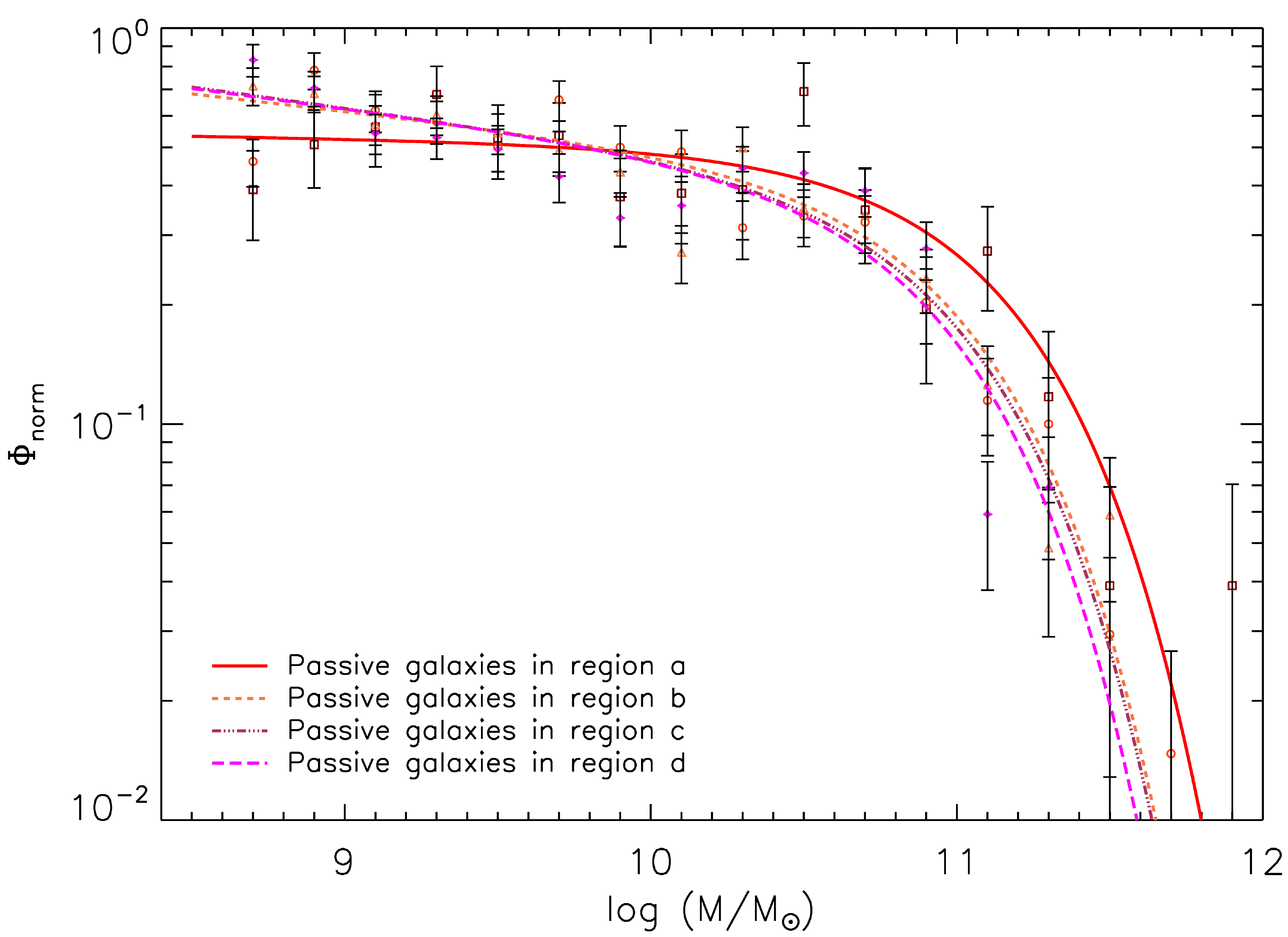}
\caption{SMFs of  passive galaxies in different density regions. The SMFs are normalized to the the total number of galaxies in each sample. Red circles, orange triangles, maroon squares and magenta diamonds  are number counts in Region (a) to (d).}
  \label{f:smf_p}
  \end{figure}

\begin{figure}[ht]
\includegraphics[width=\columnwidth]{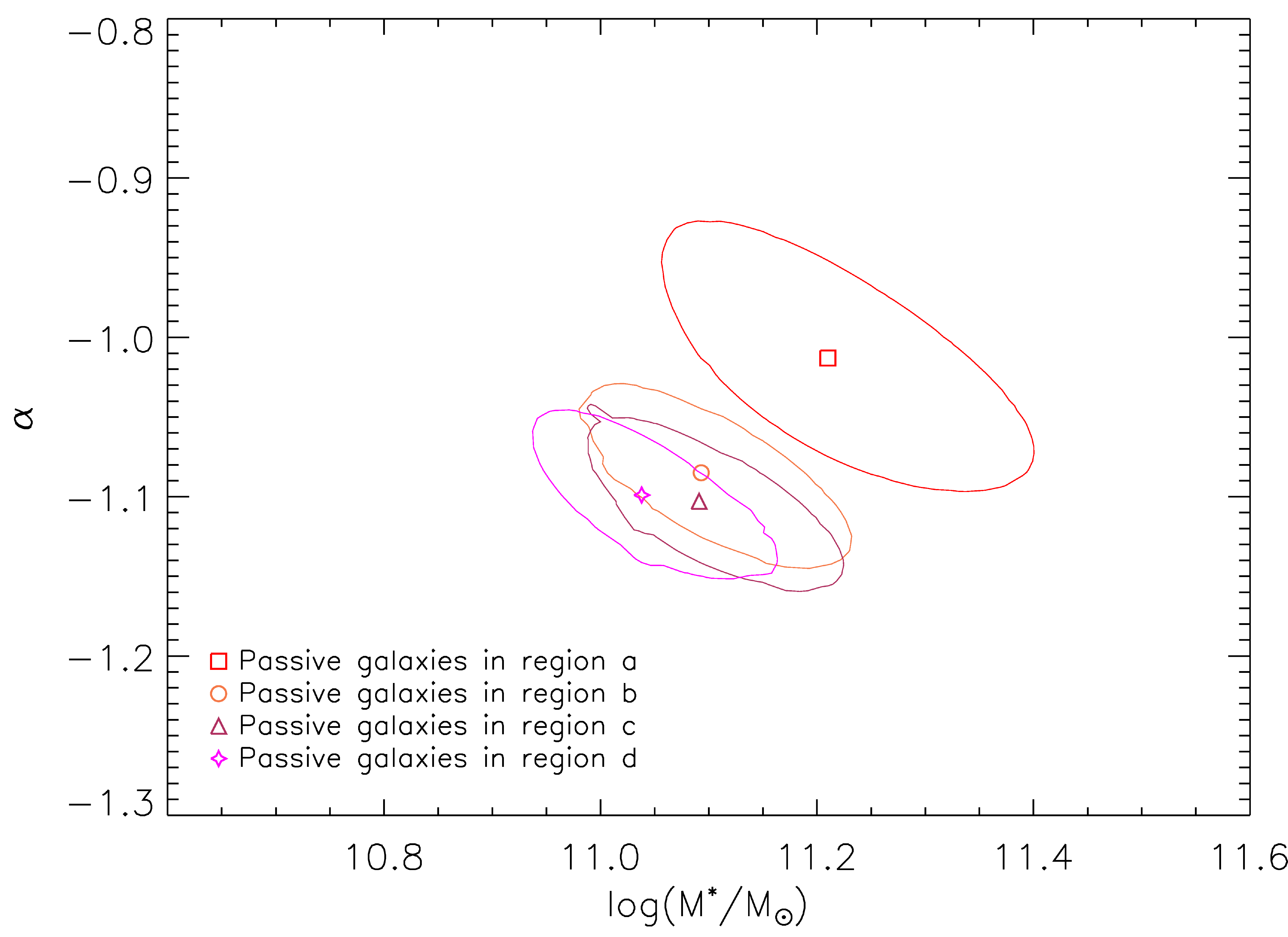}
  \caption{As in  Fig.~\ref{f:sbf}: best-fit Schechter parameters $\mstar$ and
      $\alpha$ with their 1~$\sigma$ contour in different density regions of the cluster.}
  \label{f:c_p}
  \end{figure}

From Fig. \ref{f:dens} one can see that regions identified by $\mathrm{\Sigma}$ are not spherically symmetric. Another peculiar characteristic is that Region (a) is not centered on the BCG and that  Regions (b) and (c) incorporate also a group in the N-NE direction. 
 \cite{mercurio2003_2} found that the  center of the cluster was coincident with the position of the BCG. However, they used a sample of brighter spectroscopically confirmed member galaxies. Their magnitude limit (\textit{R} < 22 mag) roughly corresponds to a mass limit $\mathrm{>\, 10^{9.5}\, \msun}$ (see Fig.~\ref{f:Mrc}). If we limit our analysis to this sample of galaxies the center is coincident with the position of the BCG and the region at high density in the NW direction disappears. In other terms the morphological appearance of the cluster changes according to the limiting magnitude of the sample.\\
The group in the N-NE direction contains very few cluster spectroscopic members. For this reason, after we checked that its presence does not affect the SMF of galaxies in different regions, we removed it from our analysis. \\
Fig.~\ref{f:smf_p} shows the SMF of passive galaxies in Regions (a) to (d), and their best-fit parameters are reported in Fig.~\ref{f:c_p} and in Table~\ref{t:sbf}. From Figs.~\ref{f:smf_p} and 
~\ref{f:c_p} one can see that the SMF of passive galaxies depends on the local density. The environment has two effects  on the SMF of passive galaxies in the densest region of the cluster: there is a higher number of high-mass galaxies than in less dense regions (that corresponds to a higher value of $\mstar$) but there is also a drop at the low end of the SMF. \\
We applied the K-S test to the $\mste$ distributions of separately SF and passive galaxies in adjacent regions. According to the results of the K-S test, in Regions (a) to (d) the $\mste$ distributions of both SF and passive galaxies are not statistically different. However, this can be related to the fact that low-mass galaxies statistically dominate the samples. If we repeat the test restricting it to the galaxies with $\mste \mathrm{\, > 10^{9.5}\, \msun}$, we find a significant difference between the mass distribution of passive galaxies in Regions (a) and (b).

\subsection{Double Schecter} 
\label{ss:ds}
Even if the single Schechter is statistically an acceptable fit to all the SMFs presented in this paper, the dip existing in the number counts of passive galaxies in all environment and the upturn in the number counts at the low-mass end visible in the low density regions seem to indicate that perhaps a double Schechter function is more suitable to describe the SMF of this galaxy population. For this reason we fit a double Schechter function to the SMF of passive galaxies in all the environments. The double Schechter function used has four free parameters:
\begin{eqnarray}
   \mathrm{\Phi (\log \textit{M}) = \ln(10) \, \fis \, \times} & \nonumber \\
   \mathrm{\left[\left(\frac{\textit{M}}{\mstar}\right)^{1+\alpha}
    + f_2
   \left(\frac{\textit{M}}{\mstar}\right)^{1+\alpha_2}
   \right] \times \exp\left(-\frac{\textit{M}}{\mstar}\right) \, d(\log \textit{M}).} &
\label{eq:double}
\end{eqnarray}

\begin{figure}
\includegraphics[width=\columnwidth]{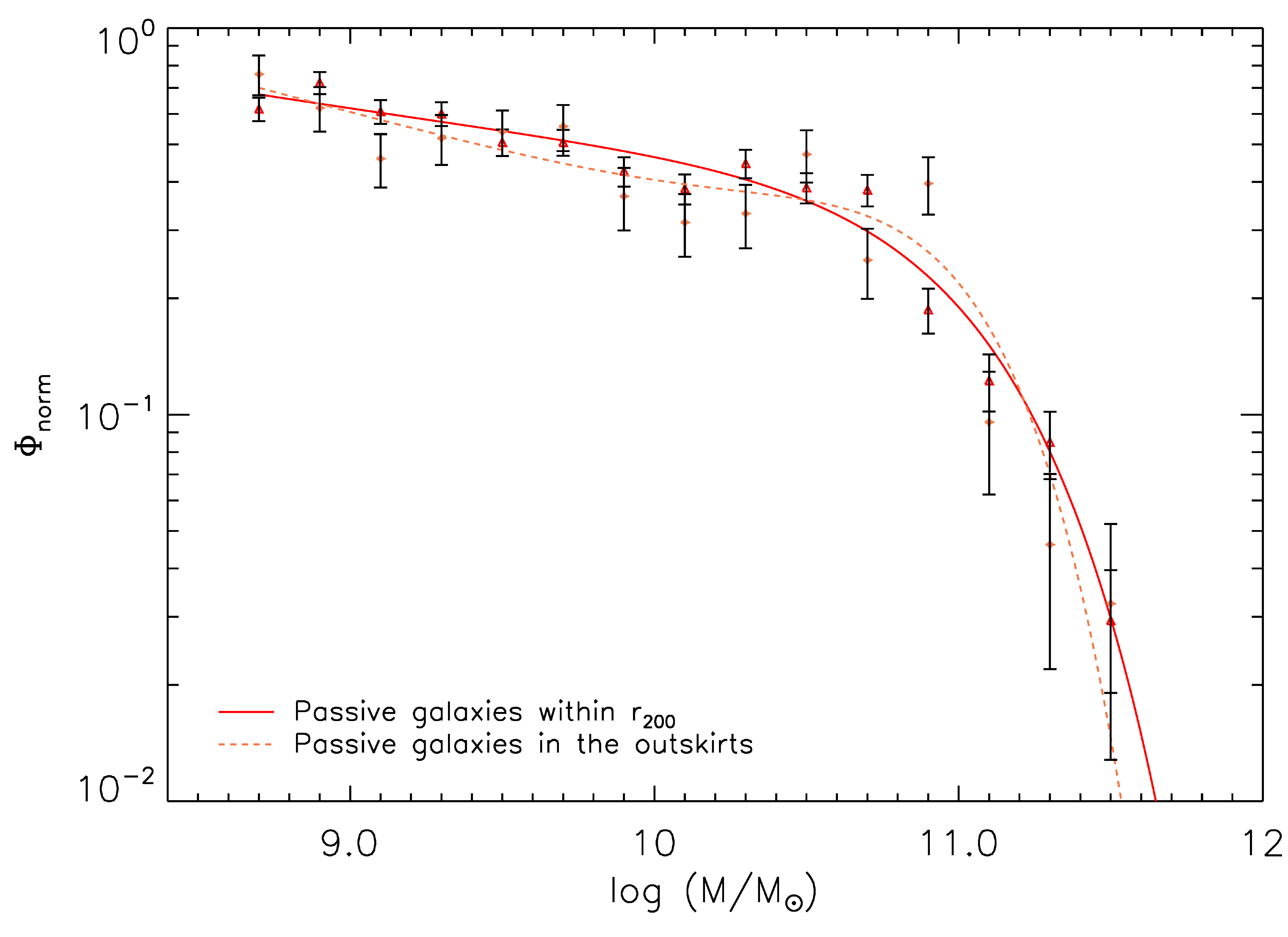}
\caption{SMF of  passive cluster member galaxies within and outside $\rtwo$. The red solid and orange dashed lines are the best-fit double Schechter functions. The points and the relative errorbars have the same meaning as in Fig.~\ref{f:smf_in_out} (right panel)}.
\label{f:ds1}
\end{figure}

\begin{figure}
\includegraphics[width=\columnwidth]{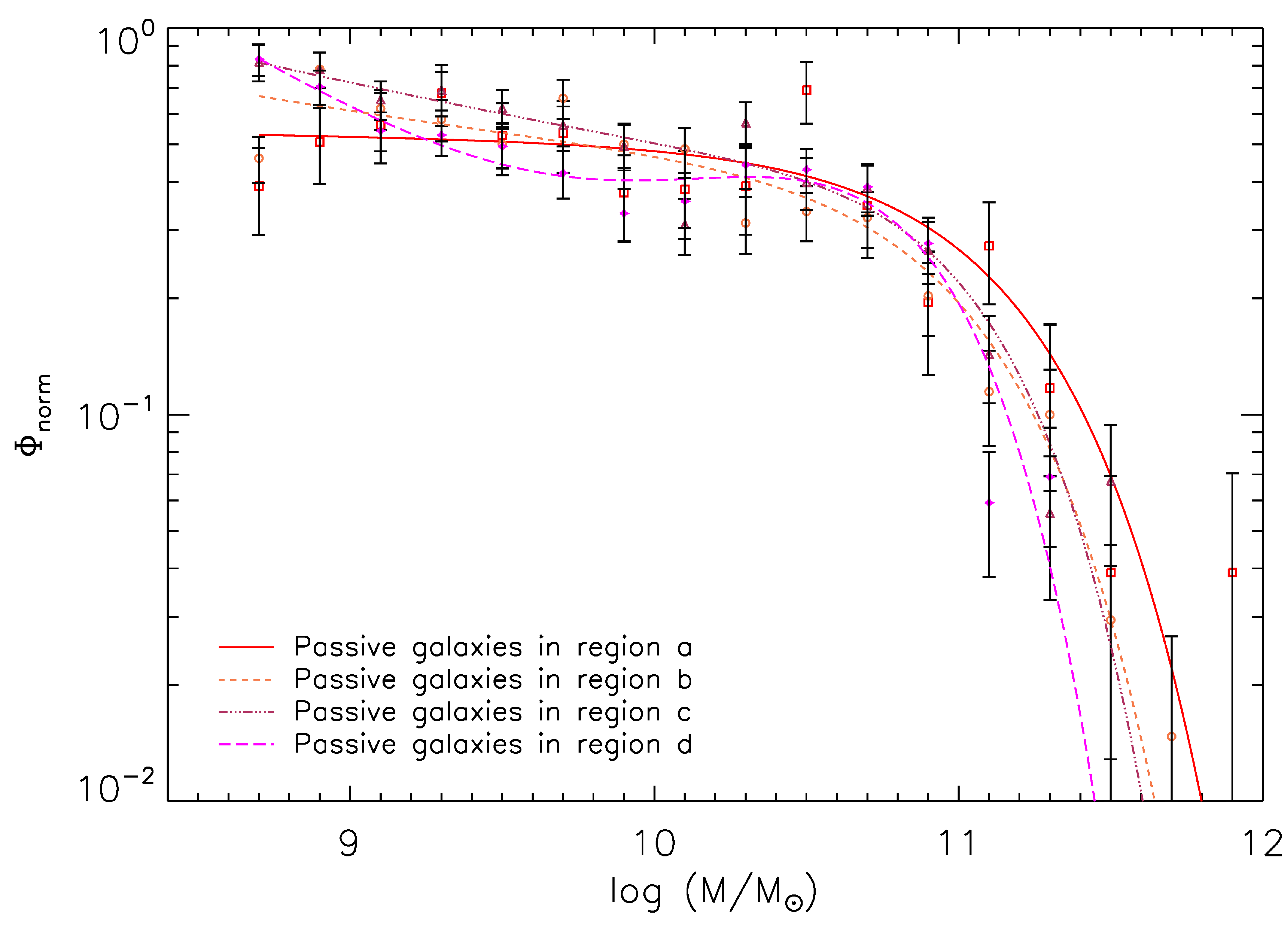}
\caption{SMFs of  passive cluster member galaxies in Regions (a) to (d). The lines are the best-fit double Schechter functions. The points and the relative errorbars have the same meaning as in Fig.~\ref{f:smf_p}.}
\label{f:ds2}
\end{figure}

Compared to the single Schechter function there are two additional free parameters: the slope of the second component $\alpha_2$, and the ratio between the normalizations of the two  functions, $\mathrm{f_2=\Phi^*_2/\Phi^*}$. We consider the same $\mstar$ for the two functions to minimize the number of free parameters and to follow the approach already used for the study of the SMF of field galaxies \citep{mortlock2015, baldry2012}. In Fig. ~\ref{f:ds1}, we show the double Schechter fit to the SMF of passive galaxies within and outside the virial radius of the cluster, while in Fig. ~\ref{f:ds2} we show the double Schechter fit in Regions (a) to (d). The best-fit values of the parameters are listed in Tab.~\ref{t:ds}. From both Fig. ~\ref{f:ds2}  and Tab.~\ref{t:ds}, one can note that in the most dense region of the cluster the slope of the two Schechter functions are statistically indistinguishable, meaning that a double Schechter function does not provide an improved fit with respect to a single one. 
The evidence of two populations of passive galaxies becomes stronger when moving to outer and less dense regions of the cluster. This is also shown by the increase of the difference between the two slopes of the Schechter functions. 
The second component in the SMF of passive galaxies is hypothetized in the model of \cite{peng2010}, as a result of the environmental quenching of SF galaxies. The fact that we observe a second component stronger in outer regions could mean that the environmental processes responsible for the quenching are most effective in external regions of the cluster or that in the center these mechanisms are so strong to completely destroy this population of galaxies.    

\begin{table*}[ht]
\centering
\caption{Best-fit parameters of the double Schechter function of passive galaxies in all environments.}
\label{t:ds}
\begin{tabular}{lcccccc}
 \hline
Environment  & $\alpha$ & $\mathrm{log(\mstar/\msun)}$ &
$f_{2}$ & $\alpha_{2}$\\
\hline
Passive & -0.61 $\mathrm{\pm}$ 0.25 & 10.96 $\mathrm{\pm}$ 0.08 & 0.76 $\mathrm{\pm}$ 0.25 & -1.18 $\mathrm{\pm}$ 0.23 \\
$\mathrm{r \, \leq \, \rtwo}$ & -0.93 $\mathrm{\pm}$ 0.27 & 11.08 $\mathrm{\pm}$ 0.07 & 0.76$\mathrm{\pm}$ 0.33  & -1.18 $\mathrm{\pm}$ 0.28 \\
$\mathrm{r \, > \, \rtwo}$  & -0.05 $\mathrm{\pm}$ 0.34 & 10.77 $\mathrm{\pm}$ 0.10 & 0.44 $\mathrm{\pm}$ 0.25 & -1.21 $\mathrm{\pm}$ 0.10\\
Region (a)  & -1.01  $\mathrm{\pm}$ 0.14 & 11.21 $\mathrm{\pm}$ 0.12 & 0.65 $\mathrm{\pm}$ 0.41 & -1.01 $\mathrm{\pm}$ 0.22 \\
Region (b)  & -0.94  $\mathrm{\pm}$ 0.08 & 11.06 $\mathrm{\pm}$ 0.10 & 0.34 $\mathrm{\pm}$ 0.35 & -1.24 $\mathrm{\pm}$ 0.26 \\
Region (c)  & -0.41  $\mathrm{\pm}$ 0.35 & 10.93 $\mathrm{\pm}$ 0.10 & 0.81 $\mathrm{\pm}$ 0.32 & -1.19 $\mathrm{\pm}$ 0.09 \\
Region (d)  & -0.34  $\mathrm{\pm}$ 0.36& 10.70 $\mathrm{\pm}$ 0.11 & 0.10 $\mathrm{\pm}$ 0.23 & -1.48 $\mathrm{\pm}$ 0.29 \\
\hline
\end{tabular}
\end{table*}

\begin{figure}
\includegraphics[width=\columnwidth]{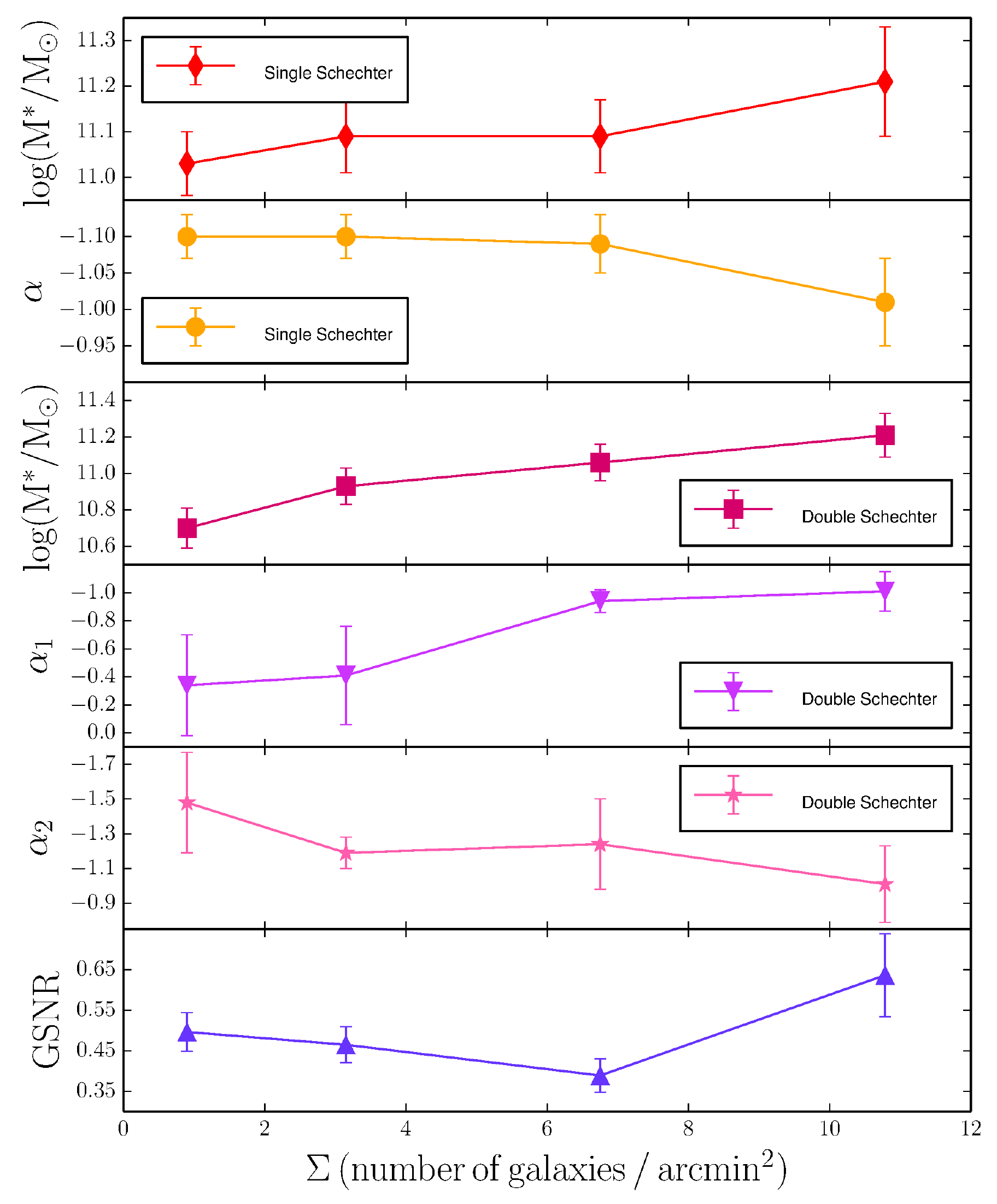}
\caption{From top to bottom: best-fit parameters of the single Schechter (panels 1-2) and of the double Schechter functions (panels 3-5) for passive galaxies and the GSNR (panel 6) as a function of the cluster local number density. Regions (d) to (a) from left to right. }
\label{f:a_m_sigma}
\end{figure}

In Fig.~\ref{f:a_m_sigma}, we show the parameters of the SMF of passive galaxies as a function of the local density of the cluster. We also show the ratio between the number of giant and subgiant galaxies (GSNR) as a function of the environment. We divide the giant and subgiant galaxies using a mass threshold of $\mathrm{10^{10.0}\msun}$, i.e. the mass value at which the SMFs of passive galaxies in region (a) and (b) intersect.
The trend in the GSNR shows that in the Region (a) the ratio between the number of giant and subgiant galaxies is higher than in less dense regions. This trend is consistent with the one we found in M1206.   

\section{The Intra-cluster light}
\label{s:icl}
In \cite{annunziatella2014}, we found that the SMF of passive galaxies in the innermost and most dense region of the cluster M1206 declines at the low-mass end. We suggested that this steeper drop could be explained by the tidal stripping of subgiant galaxies ($\mathrm{\mste \, < \, 10^{10.5} \msun }$). In our scenario, the material stripped by these galaxies could contribute to the ICL. To check if this hypothesis holds also in this cluster  we need to obtain an ICL map and an estimate of the ICL color and mass. \\
The "intra-cluster light"  is a diffuse component which originates from stars bound to the cluster potential that are tidally stripped from the outer regions of galaxies that interact with the BCG or other cluster members (\citealt{murante+07}, \citealt{contini+14}). Some properties of the ICL, such as the radial profile, the color, and presence of substructure are strictly linked to the accretion history and dynamical evolution of galaxy clusters. For example, the presence of tidal arcs can give information on the assembly history of the cluster. If the ICL is formed before the cluster has undergone virialization, the bulk of the ICL will be smooth with few faint tidal arcs. According to theoretical models these features are the result of recent interaction among galaxies with the cluster potential (\citealt{rudick2009}).
The color can also give some hints on which are the main progenitor galaxies of the ICL. 
Even if the ICL can shed light on many mechanisms occurring in galaxy clusters, it has proved very hard to study it due to its very faint surface brightness, i.e. $\mathrm{\sim \, 1\%}$ of the sky brightness. 
Furthermore, the detection of the ICL at high redshift is made even more difficult by
 the surface brightness dimming, that scales with redshift as $\mathrm{(1\,+\,\textit{z})^4}$.\\
To obtain the ICL map we follow the method developed by \citet{Presotto+14}. In the following sections, we summarize the main aspects of this method. 

\subsection{ICL detection method}
\label{ss:icl_m}

There is no consensus on the method for the detection of the ICL. Ideally, one should subtract the light contribution from all galaxies in the cluster including the BCG. However, it is not always possible to perfectly fit the light distribution of each galaxy. For this reason, in many works the galaxies are masked down to a certain limit in surface brightness. The approach we use is a hybrid method which first fits a S\'ersic profile to each galaxy and subtracts the best-fit model whenever it is possible, then it masks all the pixels with high residuals. We apply this method to analyze the $R_C$ image of the cluster. Since the region interested by the ICL is  $\mathrm{\sim 500\, kpc}$ around the BCG, we limit our analysis to a squared image of size $\mathrm{\sim 1 \, Mpc}$ centered on the BCG.
We use an automated procedure which relies on the GALAPAGOS software (\citealt{barden2012}). The main goal of GALAPAGOS is to obtain a S\'ersic fit for all the sources detected in the image by SExtractor \citep{bertin1996}. The fits are performed by the GALFIT code (\citealt{peng2010_gal}). Detailed explanations of how GALAPAGOS works is given in \citet{barden2012}. Here, we briefly describe only the steps that are crucial for our analysis. \\
The detection of the sources is performed by SExtractor in two-step runs: one so called "cold"  tuned to properly detect and deblend only bright sources, and another called "hot",  focused on the detection of faint ones. The two detections are then combined, by rejecting the faint sources whose position falls inside the cold source as determined
by the Kron ellipse (\citealt{kron1980}). The input parameters are chosen to detect in the "hot" mode sources with $\mathrm{S/N \, > \, 1\sigma_{sky}}$. For each detected source, GALAPAGOS creates a stamp image centered on the source position large enough to include neighbor galaxies. Then, GALFIT simultaneously fits all the sources in each stamp. The starting parameters for the fits are the SExtractor estimate of source position (\textit{X\_IMAGE, Y\_IMAGE}), the source total magnitude (\textit{MAG\_BEST}), its effective radius (function of \textit{FLUX\_RADIUS}), and its position angle (\textit{THETA\_IMAGE}). The 
output parameters from the fits and those obtained by SExtractor are then combined in a final catalog.  
Then we use the GALtoICL procedure presented in \citet{Presotto+14}. This procedure merges the stamps created by GALAPAGOS to form a global model image which is later subtracted to the original one to obtain a residual image called BCG+ICL map. The sources that have high residuals are identified by comparing the distribution of the residuals with the distribution of the pixel values of a sky region. The pixels whose values deviate at different $\mathtt{\sigma}$ levels from the sky ($\mathtt{1,\, 2, \, 3, \, 4, \, 5\, \sigma_{sky}}$) are flagged and associated to the corresponding source. These pixels can be masked or one can decide to manually re-fit the sources to which they belong. Before the BCG+ICL map is created one can also decide to manually add masks (for example in case of stellar spikes).\\
\begin{figure*}[ht]
\includegraphics[width=0.41\linewidth, bb=0.0cm 0cm 10cm 8cm]{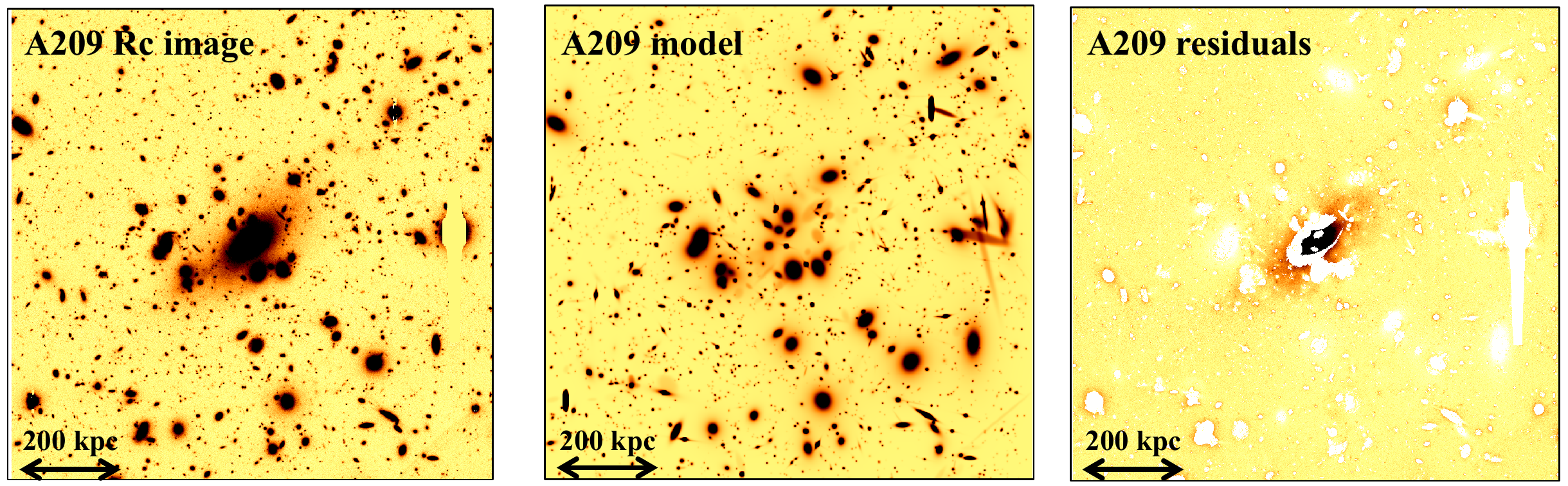}
\caption{ SuprimeCam image in the $R_C$ band of A209 (first panel), the global model image (second panel) and the BCG+ICL map with pixels at $\mathrm{1 \, \sigma_{sky}} $ level (third panel). The images are $\sim$ 60 arcmin for each side.}
\label{ICL_map}
\end{figure*}  
In Fig. ~\ref{ICL_map}, we show the $R_C$ image of A209 (first panel), the global model image (second panel) and the BCG+ICL map with pixels at $\mathrm{1 \, \sigma_{sky}} $ level (third panel). We can check if the subtraction of the light from the sources was done in an efficient way by counting the percentage of masked pixels. In our case this percentage varies from 7\% to 10\%  if we mask at $\mathrm{5 \, \sigma_{sky}} $ or $\mathrm{1 \, \sigma_{sky}} $.  \\

\subsection{ICL properties}
\label{ss:icl_p}

We estimate the $\mathrm{B\, -\, R_C}$ color of the BCG+ICL map. To do this we take as a benchmark the global model in the $R_C$ band. From this we obtain a model for the B band by changing the zeropoint and fitting only the magnitudes of the objects. Then, we subtract this model from the B-band image of the cluster to get the B band BCG+ICL map. At this point, we degrade the BCG+ICL map in the $R_C$ band to the PSF of the B band by convolving it with a Gaussian function whose $\sigma$ is the difference between the two PSFs and from this we subtract the B band BCG+ICL map.
The $\mathrm{B\, -\, R_C}$ color of the BCG+ICL is $\mathrm{\sim}$ 2.1 mag at the position of the BCG while it drops to 1.2 mag at a distance from the BCG of $\mathrm{\sim 110\, kpc}$.  In this radial range, the color of the BCG+ICL remains in the color range of the passive cluster members (see Fig. ~\ref{f:Mrc}). \\
The surface brightness profile of BCG+ICL deviates from a single S\'ersic profile (\citealt{Presotto+14}). For this reason we fit the BCG+ICL light profile with a sum of multiple components. Moustakas et al. (in prep.) model the surface brightness profile of the BCGs of each CLASH cluster. They find that, contrary to the other BCGs in the sample, the light profile of the BCG of A209 deviates from a single S\'ersic profile. They perform the fits in all HST bandpasses and assume an universal value of $n \mathrm{\, and \, } \re$. 
We choose to fit the BCG+ICL profile with three components: two components for the BCGs with the effective radius ($\re$) and the S\'ersic index fixed to the values found in Moustakas et al., and another one for the ICL. The best-fit parameters for the three component are listed in Tab ~\ref{t:icl}.  The position angle of the three components are consistent within the uncertainties. The reduced $\mathrm{\chi^2}$ of the fit is 1.535.\\
\begin{table}[h]
\centering
\begin{tabular}{cccc}
\hline 
\hline
Component & $\re$ & \textit{n} &  $\mathrm{mag_{tot}}$ \\
                 & ($\mathrm{h_{70}^{-1}\, kpc}$ ) & & (AB mag)  \\
\hline
1 (BCG) & 0.86 & 0.5 & 20.07 \\
2 (BCG) & 44.28 & 3.07 & 16.15 \\
3 (ICL)  & 120 & 1.51 & 16.57 \\
\hline
\end{tabular}
\caption{Best-fit parameters of the three component of the S\'ersic fit of the BCG+ICL. }
\label{t:icl}
\end{table}
We used the magnitude value obtained from the fit to estimate the ICL mass.
We perform a bi-weighted fit of the relation between 
the stellar masses of red cluster member galaxies which are in the subsample image analyzed with GALAPAGOS and their total $R_C$ magnitude provided by GALFIT:
\begin{equation}
\centering
\mathrm{log(\textit{M}/M_{\odot} )\, = \, (18.41 \pm 0.07) - (0.41 \pm 0.02)\times \textit R_{C_{tot mag}}.}
\label{e:rcMfit}
\end{equation}
Using this relation with the total $R_C$ magnitudes obtained from the multicomponent S\'ersic fit, we estimate $\mathrm{M_{ICL} \, =\, (2.9 \pm 0.7 \,) \times \, 10^{11} \, \msun }$ and $\mathrm{M_{BCG} \, =\, (6.2 \pm 1.2 \,) \times \, 10^{11} \, \msun }$. The mass of the BCG obtained in this way is consistent with the mass obtained applying MAGPHYS, i.e. $\mathrm{6.2 \pm 1.2 \, \times \, 10^{11} \, \msun }$.\\
To check if the ICL can form from tidal stripping of subgiants galaxies (i.e. galaxies with $\mathrm{\mste <  10^{10.0} \, \msun}$),  
we have to evaluate the missing mass from subgiant galaxies in Region (a). The missing mass from subgiant galaxies is determined with the following equation:
\begin{equation}
\mathrm{\Delta M_{sub}} \equiv \int_{10^{8.6}\, M_{\odot}}^{10^{10.0}\, M_{\odot}} \mathrm{\textit{M}\, \Phi_{a}^{*}(\textit{M})\,  d{\textit{M}} \, - \, } \int_{10^{8.6}\, M_{\odot}}^{10^{10.0}\, M_{\odot}} M\mathrm{\, \Phi_{a}(\textit{M})\,  d\textit{M}, }
\label{e:msub}
\end{equation}
where $\mathrm{\Phi_{a}}$ is the SMF of passive galaxies in Region (a) not normalized, while $\mathrm{\Phi_{a}^*}$ is the SMF of passive galaxies if we substitute the value of the slope with the best-fit value obtained for the SMF in Region (b).  The integral is done between the completeness mass limit and the mass at which the SMFs of passive galaxies in Regions (a) and (b) intersect each other ($\mathrm{\sim 10^{10.0} \, \msun}$).

The upper and lower error on $\mathrm{\Delta M_{sub}}$ are estimated by repeating the evaluation of Eq.~\ref{e:msub} and fixing the slope to $\mathrm{\alpha \, \pm \, \delta \alpha}$. The value of $\mathrm{\Delta M_{sub}}$ is $\mathrm{1.4^{+0.5}_{-0.7} \, \
\times \, 10^{11}\, \msun}$. The value of $\mathrm{\Delta M_{sub}}$ is formally in agreement with $\mathrm{M_{ICL}}$ within $\mathrm{1\, \sigma}$. A similar agreement between  $\mathrm{\Delta M_{sub}}$ and $\mathrm{M_{ICL}}$ was also found in M1206.
Breaking the integral in Eq.~\ref{e:msub} we can estimate the contribution to  $\mathrm{\Delta M_{sub}}$ coming from galaxies in different mass ranges: the main contribution is given by galaxies in the mass range $\mathrm{10^{9.5}\, < \, \mste \, < 10^{10.0}\,  \msun}$ (60\%), galaxies in the mass range $\mathrm{10^{9.0}\, < \, \mste \, < 10^{9.5}\,  \msun}$ contribute for $\sim$ 30\% ,while galaxies in the range $\mathrm{10^{8.6}\, < \, \mste \, < 10^{9.0}\,  \msun}$ only for 10\%.  This suggests that the ICL could originate from stripping of stars from subgiant galaxies in the mass range $\mathrm{10^{9.0}\, < \, \mste \, < 10^{10.0}\,  \msun}$. On the contrary, this result seems to rule out that the disruption of dwarf galaxies is the main channel for the formation of ICL, in agreement with the predictions from semi-analytical models (\citealt{contini+14}). 
Our result is also consistent with recent findings from observational works. In detail, \cite{demaio2015}, use a subsample of CLASH clusters at $\mathrm{0.44\, \leq \, \textit{z} \, \leq \, 0.57}$ to determine the color gradients and total luminosity of the ICL. Their main conclusion is that the ICL originates from the tidal destruction of 0.2\textit{L*} galaxies (which corresponds to galaxies of $\mathrm{\sim \, 10^{10.0} \msun}$). In a completely independent way, \cite{longobardi2013} assess that the Planetary Nebulae in the Intra-cluster medium of Virgo cluster can be associated to galaxy progenitor of mass around four times the mass of the Large Magellanic Cloud ($\mathrm{\sim \, 10^{9.5}\, \msun}$).

\section{Orbits of passive galaxies}
\label{s:orbits}
To investigate further the evolution of galaxies in A209, we carry out the dynamical analysis of the orbits of passive galaxies.
The mass profile of A209 has been determined from gravitational lensing
by \citet{merten2014}. Parametrized with a NFW profile \citep{NFW97}, it is characterized by a concentration $c_{200}\, =\, 3.3 \pm 0.9$ and a mass $M_{200}\, =\, (1.4 \pm 0.1)\, 10^{15} \, \msun$. These values correspond to a NFW scale and virial radius of
$\rmtwo \, =\, 0.6 \pm 0.2$ Mpc and $\rtwo \mathrm{\, =\, 2.1 \pm 0.05}$ Mpc, respectively.
Given this mass profile, it is possible to invert the Jeans equation for the dynamical equilibrium \citep{BT87} to determine the orbits of different populations of cluster galaxies \citep{BM82}. The orbits are described by the velocity anisotropy profile
\begin{equation}
\br = 1 - {\sigma_\theta^2(r) + \sigma_\phi^2(r)  \over
  2\,\sigma_r^2(r)} = 1 - {\sigma_\theta^2(r) \over \sigma_r^2(r)},  
\label{e:beta}
\end{equation}
where $\sigma_\theta, \sigma_\phi$ are the two tangential components,
and $\sigma_r$ the radial component, of the velocity dispersion, and
the last equivalence is valid when both the density and the velocity
structures of the cluster are invariant under rotations about its center,
i.e. the cluster does not rotate.

We determine $\br$ by solving the equations in \citet{SS90}, with the
technique of \citet{biviano2013}. The error bars are determined by a bootstrap procedure, running
the inversion on 100 extractions from the original data set.

\begin{figure}
\includegraphics[width=\columnwidth]{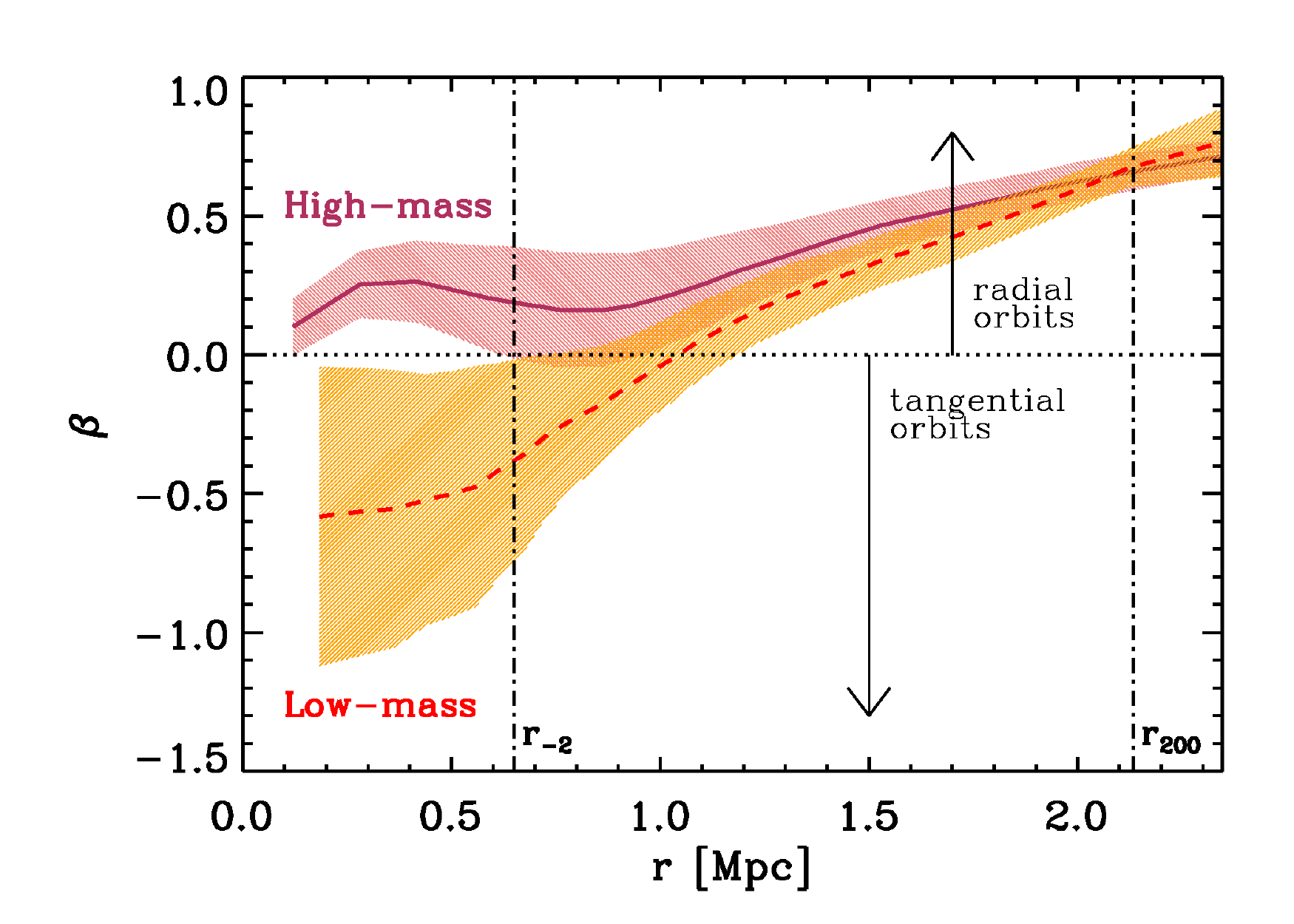}
\caption{The maroon solid and red dashed curves represent the velocity anisotropy
  profiles $\br$ of the high- and low-$\mste$ subsamples of passive
  cluster galaxies (above and below $10^{10.0} \, \msun$). The shaded
  regions represents the corresponding 68\% confidence
  levels. The vertical lines indicate the positions of $\rmtwo$
  (dashed) and $\rtwo$ (dash-dotted).}
\label{f:br}
\end{figure}

We determine $\br$ separately for two subsamples of passive galaxies,
with $\mste \mathrm{< 10^{10.0} \, \msun}$, and, respectively,
$\mste \mathrm{\geq 10^{10.0} \, \msun}$. The results are shown in
Fig.~\ref{f:br}, the red/blue curves corresponding to the
high-/low-mass subsample, respectively. The two profiles look similar
at radii $r>\rmtwo$, but they differ in the inner region.  The $\br$
of high-mass galaxies is similar to that seen in other clusters
\citep[e.g.][]{biviano2013} and also in simulated halos
\citep[e.g.][]{Munari+13}. The $\br$ profile of low-mass galaxies
indicates tangential orbits within $\rmtwo$. This might occur as a
result of selective destruction of low-mass galaxies on radial orbits
near the cluster center. Radial orbits are characterized by small
pericenters and only the more massive galaxies might be able to
survive the very hostile and dense environment of the central cluster
region. Low-mass galaxies which manage to escape destruction (e.g. by
tidal forces) near the cluster center might be those that avoid
passing very close to the cluster center by moving on slightly
tangential orbits.

\section{Mass-size relation}
\label{s:sms}
\begin{figure}
\includegraphics[width=\columnwidth]{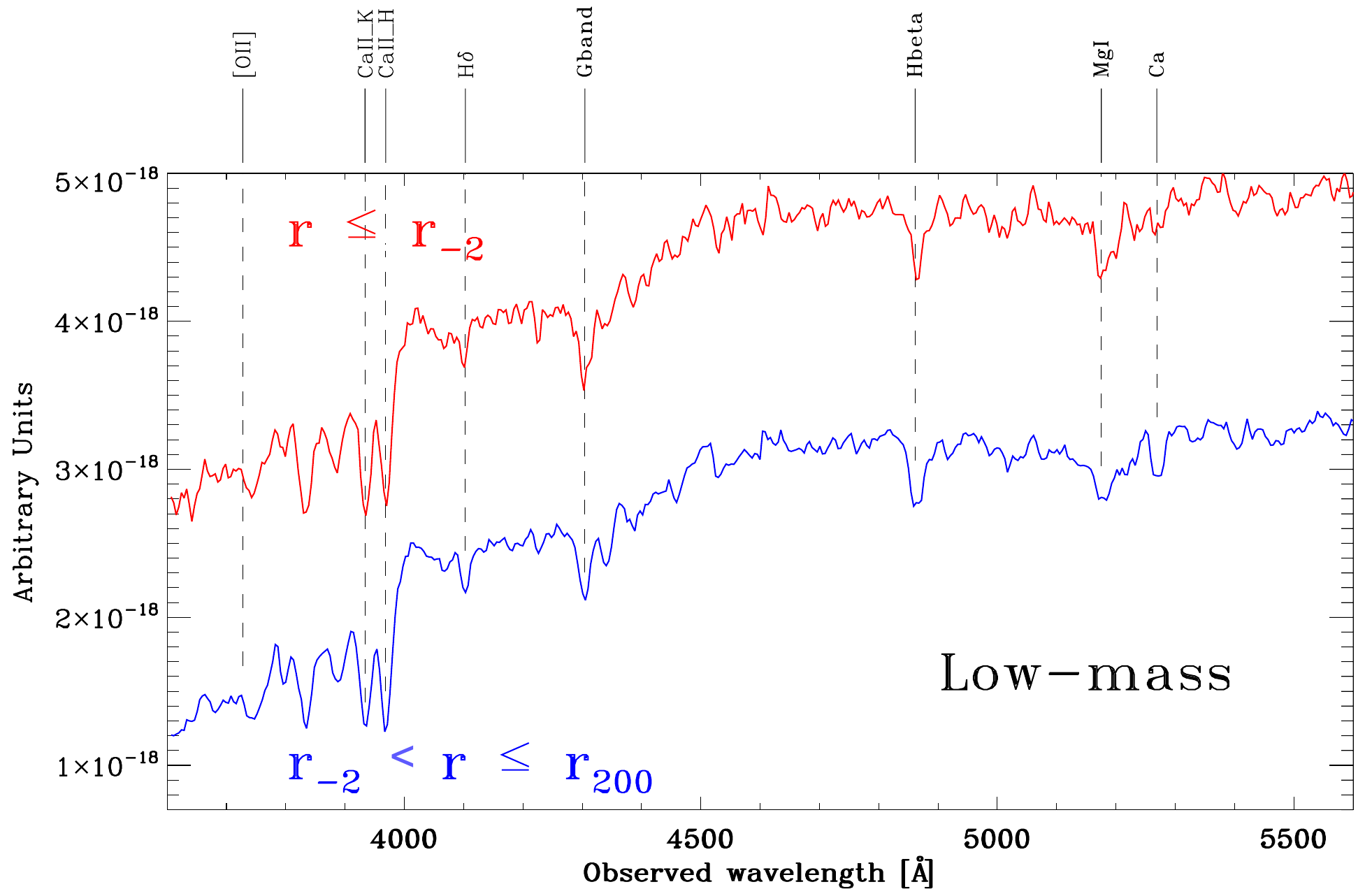}\\
\includegraphics[width=\columnwidth]{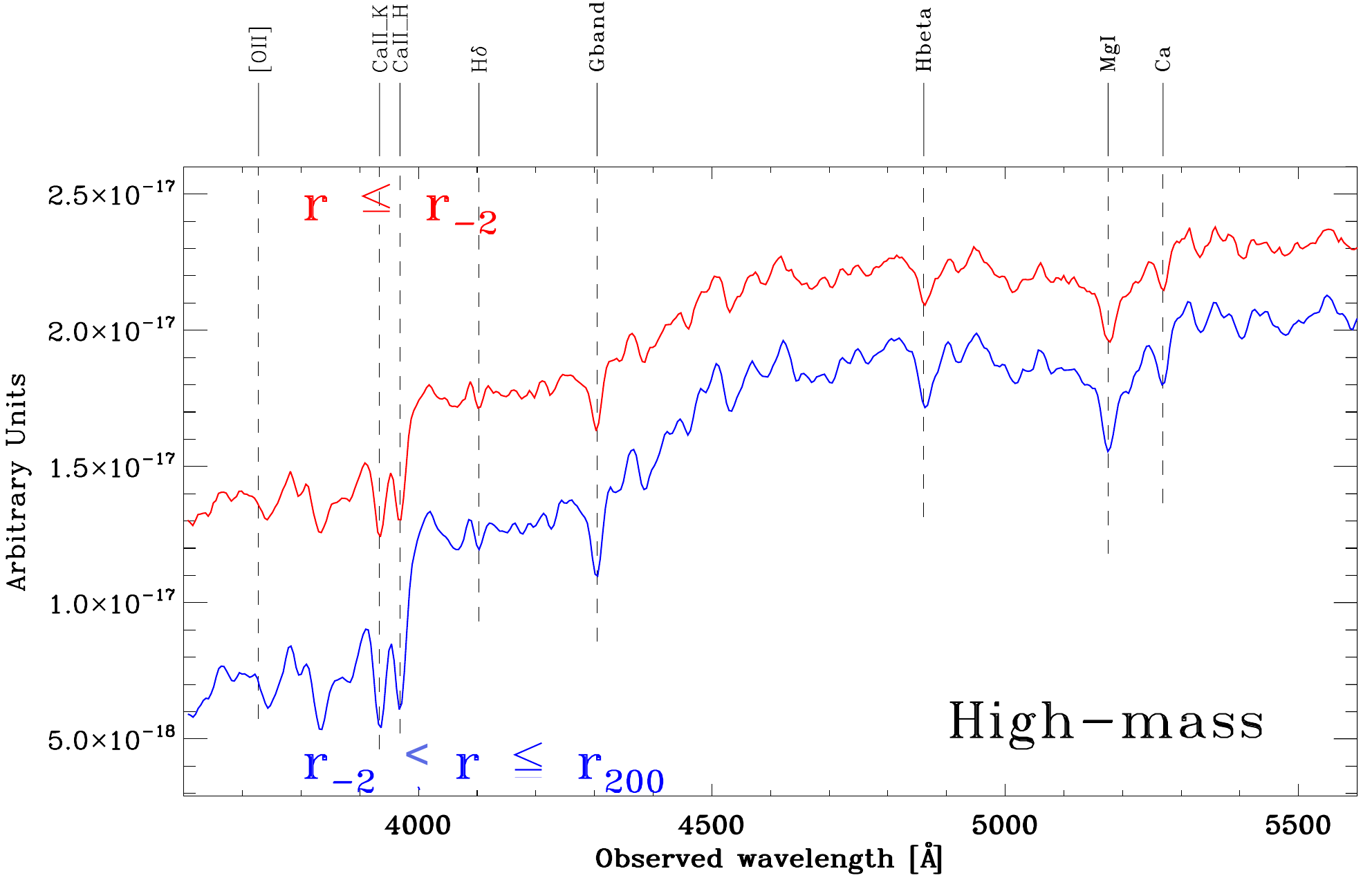}
\caption{Stacked spectra of ETGs for low-mass ($\mathrm{\leq \, 10^{10.0}\, \msun}$, upper panel) and high-mass ($\mathrm{> \, 10^{10.0}\, \msun}$, lower panel) galaxies in the internal (red) and external (blue) $\rmtwo \mathrm{\, =\, 0.65 Mpc}$.}
\label{f:ssp}
\end{figure}

\begin{figure}
\includegraphics[width=\columnwidth]{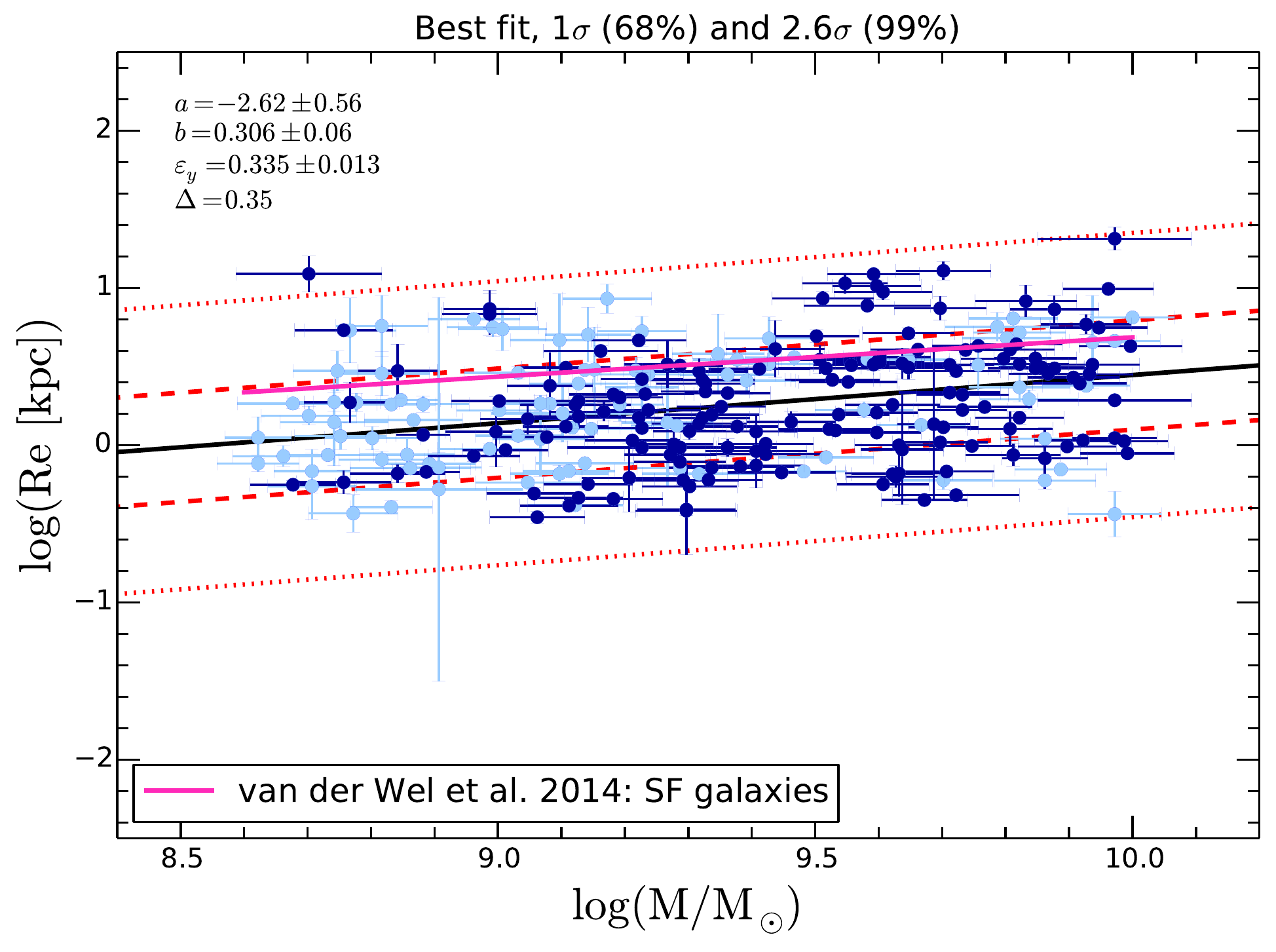}\\
\includegraphics[width=1.015\columnwidth]{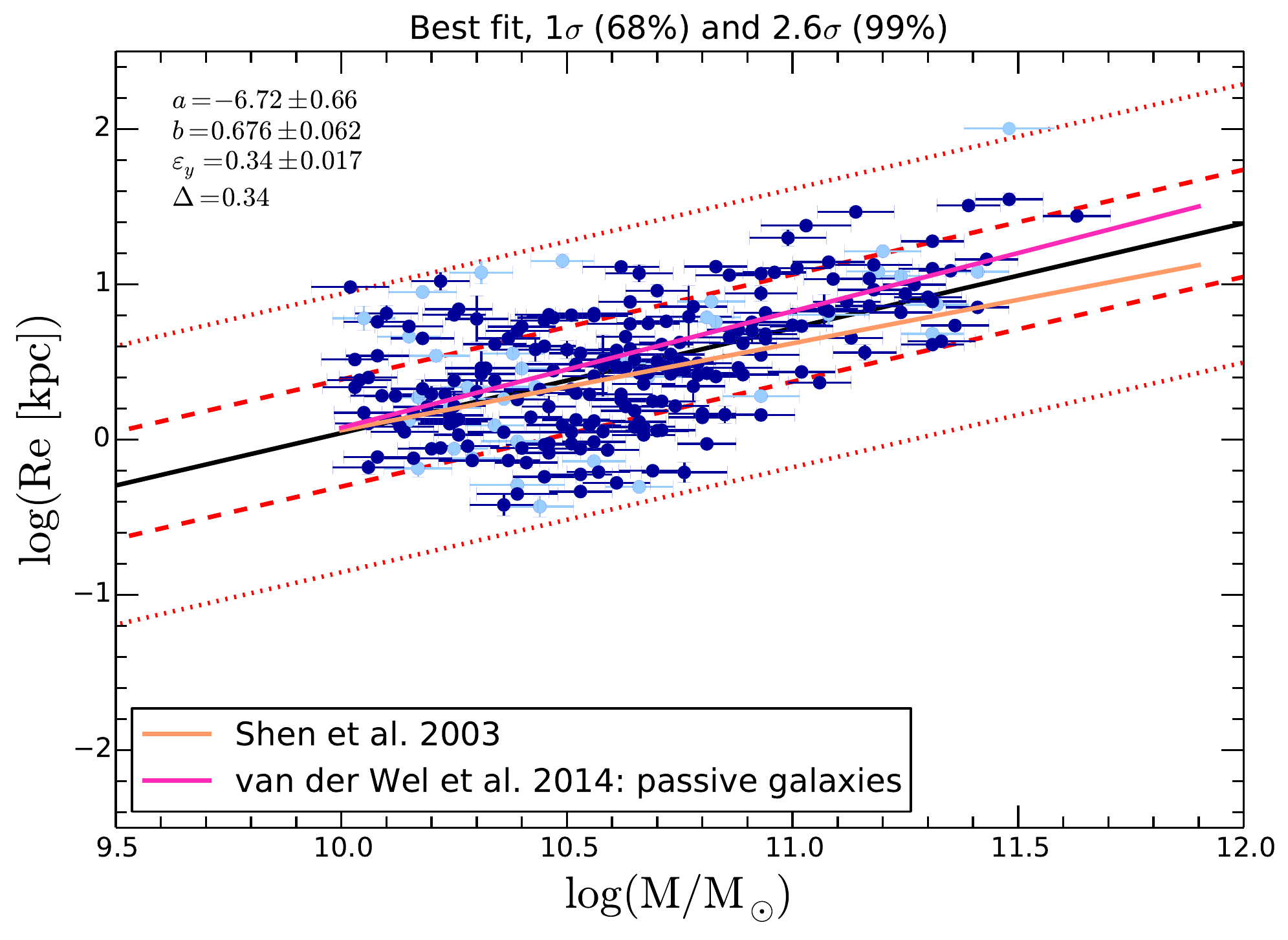}
\caption{Best-fit of the mass-size relation in Eq.~\ref{e:mre} for early-type passive low-mass galaxies  ($\mste \mathrm{\, \leq \,  10^{10}\msun}$) (upper panel) and high-mass galaxies  ($\mste \mathrm{\, > \,  10^{10}\msun}$). The best-fit values of the two parameters $a$ and $b$ with their $\mathrm{1\sigma}$ uncertainty, are reported on the two panels. Dark blue points identify spectroscopic members, while light blue point are photometric members.}
\label{f:ms}
\end{figure}

To estimate the size of each cluster member down to the mass limit of the sample, we use the GALAPAGOS software described in Sect.~\ref{ss:icl_m} applied to the SUBARU R-band image. The size of each galaxy is identified by the effective radius of the single S\'ersic fit  performed by GALAPAGOS.
The core of Abell 209 was observed with HST, as part of the CLASH program, in 16 broadband filters from UV to near-IR, using WFC3/UVIS, ACS/WFC, and WFC3/IR. Moreover two parallel fields $\sim$ 1.28 Mpc far from the cluster core were observed with ACS/WFC and WFC3/IR with filters F850LP and F160, respectively. The images were reduced using standard techniques, and then co-aligned and combined using drizzle algorithms to a pixel scale of 0.065\,arcsec (for details, see \citealt{koek2007, koek2011}).
In order to check for the reliability of the sizes we applied GALAPAGOS also on the HST image of the cluster in the F850LP band.
The median difference in the estimation of the size for those galaxies in both the SUBARU and the HST field is 0.06 arcsec (0.2 kpc). This is valid throughout the whole mass range. For this reason, we are confident in our size estimates with SUBARU even at  our mass limit ($\mathrm{10^{8.6}\msun}$), corresponding to a typical size of $\mathrm{0.3\, arcsec}$ (see Fig.~\ref{f:ms}), larger than the angular resolution in HST.\\
For the analysis of the size distribution, we use the photometric sample of galaxies defined as passive ($\sSFR \mathrm{\, <\, 10^{-10} \, yr^{-1}}$) and early-type ($n_{Sersic} \mathrm{\, > \, 2.5}$, \citealt{shen2003}).  
In order to assess the goodness of the photometric classification in passive and SF galaxies on the basis of their \sSFR and in early- and late-type on the basis of their S\'ersic index, we take advantage of the spectroscopic sample. 
We divide the galaxies of the spectroscopic sample in different spectral classes using the method described in Girardi et al. (2015). Following this classification, we divide the Emission Line Galaxies (ELGs) in weak (wELG), medium (mELG), strong (sELG) and very strong (vsELG), according to the EW[OII]. Non-emission line galaxies are divided in post-starburst, if they have EW[H$\mathrm{\delta}$]$>$3\AA (HDS), and in Passive (hereafter PAS).
In this classification we consider only galaxies with S/N > 5 for each pixel in the region of the spectrum around the H$\mathrm{\delta}$ line. At our mass limit of $\mathrm{10^{8.6} \msun}$ we have the spectroscopic classification for 741 galaxies. 
We use this spectroscopic classification in order to assess the goodness of the photometric classification in passive and SF galaxies on the basis of their \sSFR (see Sect.~\ref{s:mass}).
Among the different spectroscopic classes, 99.4\% of passive galaxies  have \sSFR < $\mathrm{10^{-10}\, yr^{-1}}$, hence identified as passive also photometrically, and 92\%  of the galaxies with equivalent width of the [OII] emission line $\mathrm{\geq}$ 15, are classified as SF on the basis of their \sSFR (> $\mathrm{10^{-10}\, yr^{-1}}$). The other ELGs are classified partially as SF and partially as passive with a mean \sSFR of $\mathrm{\sim \, 10^{-10.5}\, yr^{-1}}$ for wELGs and $\mathrm{\sim \, 10^{-10.1}\, yr^{-1}}$ for mELGs. In first approximation, we can safely use the \sSFR value to discriminate between SF and passive galaxies.
We use the spectroscopic sample also in order to test the morphological classification in early- and late-type galaxies on the basis of their S\'ersic index ($n$ > 2.5 for early-type galaxies, hereafter ETGs). Of the total sample of galaxies with spectroscopic classification, 568 are ETGs ($n$ > 2.5). Among them, 78\% are PAS galaxies, 12.5\% are wELGs and 7\% are HDS galaxies.
Following the approach already used for the orbital analysis, we divide our sample in low and high-mass galaxies with a threshold value $\mathrm{10^{10}\, \msun}$, and, then, we further splitted these two sub-samples inside and outside $\rmtwo$, to investigate possible environmental effects on the mass-size relation. In Fig.~\ref{f:ssp}, we show the stacked spectra of ETGs belonging to the four sub-samples defined above, which appear rather similar. By using the software pPXF (\citealt{cappellari2004}) and MILES models (\citealt{vazdekis2010}) we obtain that galaxies host old stellar populations with ages in the range $8.0 \pm 1.0$ Gyr in the four bins. Galaxies with $\mste \mathrm{\, > 10^{10}\,M_{\odot}}$ are consistent with solar metallicities, whereas low-mass galaxies with sub-solar metallicities.

In order to use a sample with as little contamination as possible, we choose to use for the mass-size analysis the photometric sample of passive (according to their \sSFR) and early-type galaxies.
Firstly, we fit the following linear relation to our data: 
\begin{equation}
\mathrm{log(\re\, [kpc])\, = \, a\, + b\times log(\textit{M}/\msun), }
\label{e:mre}
\end{equation}
separately for low-mass and high mass sub-samples. The fits are performed using the python code \textit{lts\_linefit.py} \citep{cappellari2013}. This code 
performs robust linear fit to data with errors in both variables.
The results of the fits are shown in Fig.\ref{f:ms}.  We compare the results obtained for the high-mass sample with the mass-size relation for galaxies in the field provided in two different works (\citealt{shen2003} and \citealt{vanderWel2014}), respectively magenta and cyan line in the lower panel of Fig.\ref{f:ms}. \cite{vanderWel2014} divide their sample of field data into early- and late-type galaxies according to their rest-frame colors. Their selection corresponds to our classification in passive and SF galaxies according to the \sSFR. 
The best-fit parameters of the mass-size relation provided in this paper are in agreement at $\mathrm{1\sigma}$ level with those in \cite{vanderWel2014} for ETGs at similar redshift. 
In this analysis we are not using circularized radii, defined as: 
\begin{equation}
R_{e_{circ}} \, = \, \re \times \sqrt{q},
\end{equation}
where $\mathrm{Re} $ is the effective radius and $q$ is the axis ratio of the single S\'ersic fit performed by GALAPAGOS, to better compare our results with those of \cite{vanderWel2014}. However, performing the fit using the circularized radii changes very little the values of the parameters ($a_{circ}\, = -8.1 \, \pm \, 0.7$, $b_{circ}\, = 0.79 \, \pm \, 0.07$). \\
We note that the slope of the relation between stellar mass and galaxy size is very different in the two range of stellar mass. In particular, for low-mass galaxies the slope of the relation is flatter than for high-mass galaxies, in agreement also with the suggestions from \cite{vanderWel2014}. Moreover, the value of the slope obtained for low-mass passive and ETGs is in agreement (within $\mathrm{1\sigma}$) with the slope of the mass-size relation presented for late-type galaxies in \cite{vanderWel2014} (see upper panel of  Fig.~\ref{f:ms}).\\
In order to investigate the effect of the environment on mass-size relation, we consider separately the galaxies inside and outside $\mathrm{\rmtwo}$ (defined in Sect.~\ref{s:orbits}), analyzing the distribution of the residuals with respect to the two relations obtained for low-mass and high-mass galaxies.
The distribution of the residuals of low-mass galaxies within $\rmtwo$ is centered on negative values. The median value of $\mathrm{\Delta \re \, \equiv \, log(\re)\,-\, log(\re{_{bf}})}$ is $\mathrm{-\, 0.12 \pm \, 0.05}$. For low-mass galaxies with $\rmtwo \mathrm{\, \leq \, r \, \leq \,} \rtwo$, instead, $\mathrm{\Delta \re \, = 0.10 \, \pm \, 0.03}$. In order to statistically assess this discrepancy we apply the K-S test to the distributions of the residuals of low-mass galaxies within and outside $\rmtwo$. The resulting K-S probability is 0.4\% which implies that the two distributions are statistically different. This result seems to indicate that at fixed stellar mass the size of low-mass galaxies in the center of the cluster are more compact than in more external regions. \\
The same test applied to the sample of high-mass galaxies confirms that the distribution of galaxy sizes are indistinguishable in the two regions. \\
The fact that the slope of the mass-size relation of low-mass passive early-type galaxies is consistent with that derived by \cite{vanderWel2014} for field star-forming galaxies is compatible with the scenario that low mass galaxies originate from quenching of late-type galaxies. The quenching mechanism could be responsible not only for the truncation of the star formation but also for a change in the morphology of the object.
The interpretation of the environmental dependence of the mass-size relation is more puzzling. According to the analysis of the orbits, low-mass galaxies tend to have tangential orbits, hence avoiding the cluster center (coincident with the position of the BCG). A possible scenario could be that when low-mass galaxies pass very close to the cluster center, they are destroyed or stripped below the mass limit. However, galaxies that are not close enough to the BCG to be destroyed still undergo tidal interactions that could reduce their size. 

\begin{figure}
\includegraphics[width=\columnwidth]{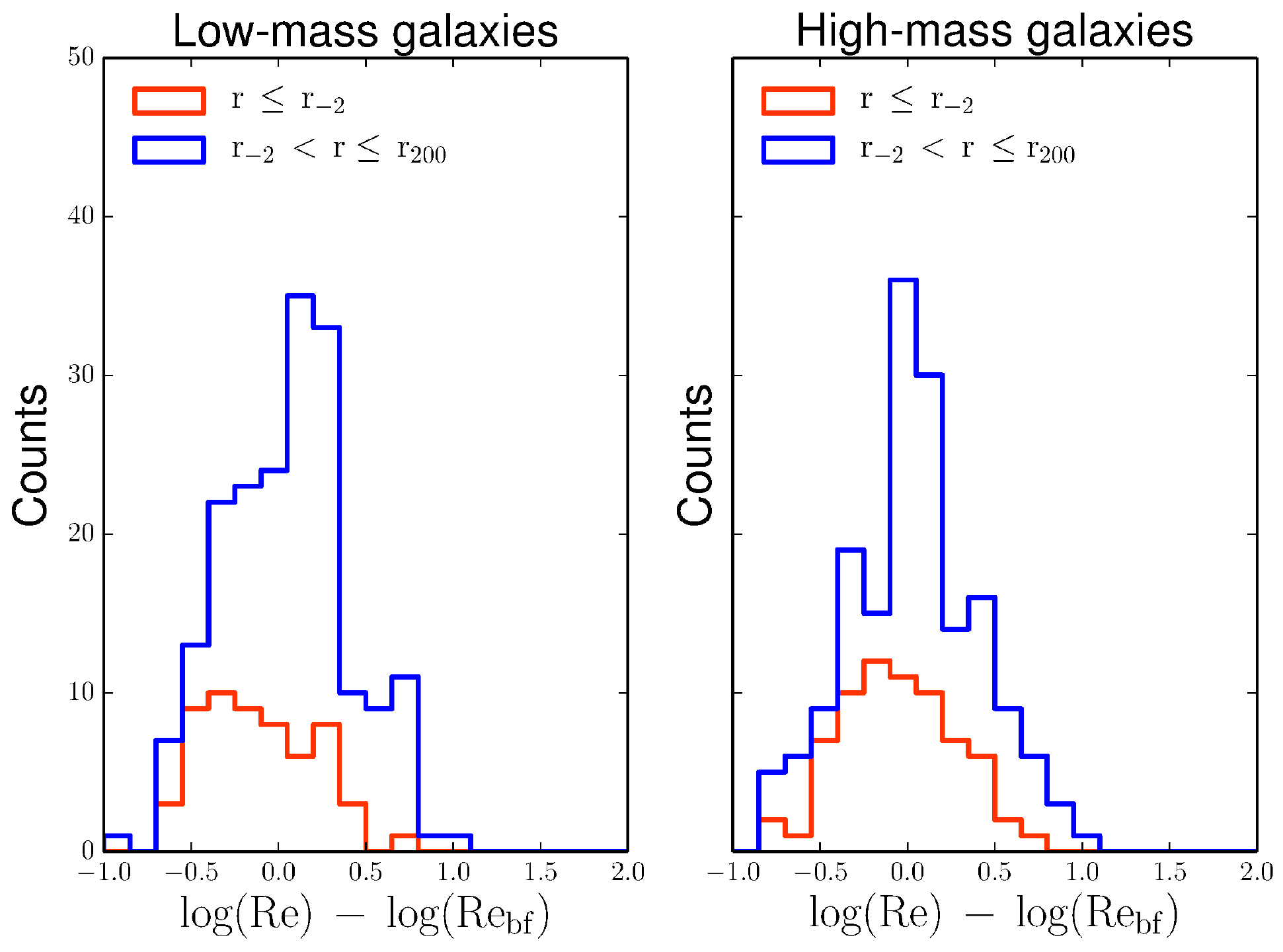}
\caption{Distribution of effective radius residuals with respect to the best-fit relation (Eq.~\ref{e:mre}) for low-mass (left panel) and high-mass (right panel) galaxies in the internal (red) and external (blue) region of the cluster.}
\label{f:res_ms}
\end{figure}

\section{The stellar mass density profile}
\label{s:smfrac}
To investigate further the environmental processes active in this cluster, we study also the number density, and the stellar and total mass density profiles. We use the sample of 1916 member galaxies with $\mste \geq
\mathrm{10^{8.6}\,M_{\odot}}$ to determine the number density (N(R)) and the stellar mass density ($\mathrm{\Sigma_{\star}(R)}$) radial profiles. 
For the determination of the number density profile, we use the same weights used for the SMF and described in Sect. ~\ref{ss:compl}, while for the determination of $\mathrm{\Sigma_{\star}(R)}$ these weights are multiplied by the galaxy stellar masses.

 \begin{figure}[ht]
\includegraphics[width=\columnwidth]{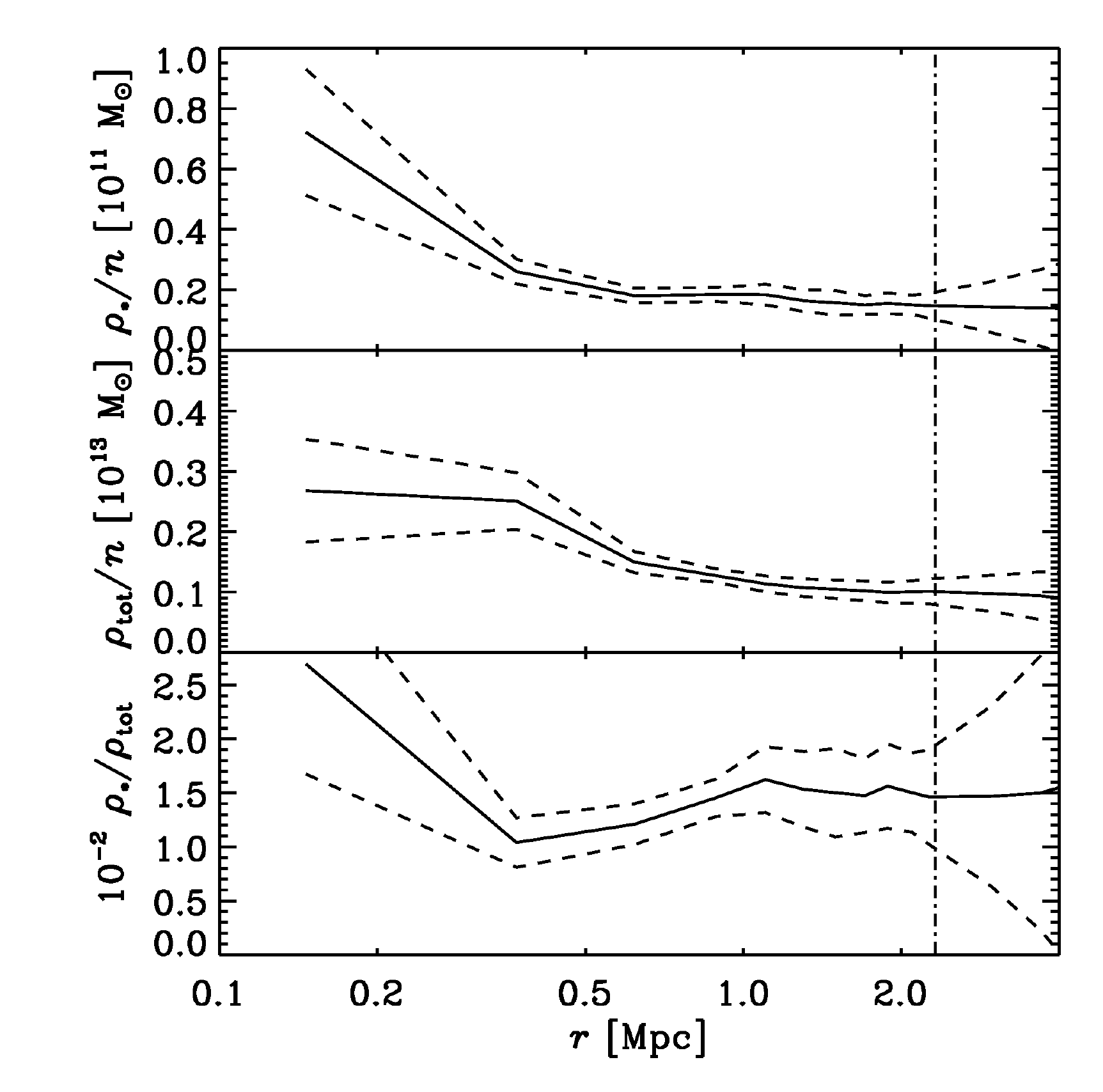}
  \caption{Top panel: the ratio of the deprojected stellar mass density and number
    density profiles. Middle panel: the ratio of the total mass
    density and number density profiles. Bottom panel: the ratio of
    the stellar mass density and total mass density profiles. Dashed
    lines indicate 1 $\sigma$ confidence regions. All densities are
    volume densities. The vertical dash-dotted line is the value of $\rtwo$. 
    }
  \label{f:profratios}
  \end{figure}

To best compare the two density profiles, we deproject them with an Abel inversion (\citealt{BT87}). This technique makes the assumption of spherical symmetry. For the inversion we use a smoothed version of the profiles obtained with the LOWESS technique \citep[e.g.][]{Gebhardt+94}, where the extrapolation to infinity is done following \citet[][eq. 10]{biviano2013}. 
The first panel of Fig.~\ref{f:profratios} shows the ratio between the deprojected mass density profile and the number density profile. The 1$\sigma$ confidence levels are obtained by propagation of errors and are represented by the dashed line assuming that the errors on the deprojected profiles are the same as for the projected profiles. The ratio shows a significant peak in the central region of the cluster. This is a striking evidence of a mass segregation effect, meaning that in the central region of the cluster the mean mass for galaxy is higher than at larger radii.
 This is due to the presence of a dozen galaxies, all with masses above $\mathrm{10^{10}\, \msun}$, within $\mathrm{0.1\, Mpc}$ from the BCG and is in agreement with the environmental dependence of the SMF.\\
An indication of the processes that are responsible for this mass segregation can be obtained comparing the mass density profile and the number density profile to the total density profile. For the total density profile ($\rho_{tot}$), we consider the estimate obtained from the gravitational lensing analysis in \citet{merten2014}. The second and third panels of Fig.~\ref{f:profratios} show the ratios  $\rho_{tot}/n$ and
$\rho_{\star}/\rho_{tot}$ as function of the 3D clustercentric distance $r$. The total mass density is less concentrated than the distribution of galaxies, in agreement with what we found for M1206, and significantly less concentrated than the stellar mass density. The ratio between stellar and total mass density profiles indicates that the mass segregation in the center of A209 is due to dynamical friction. In fact, this process transfers kinetic energy from baryons to dark matter, and this translates into a flattening of the dark matter density profile \citep[e.g. ][]{el-zant2001, DelPopolo12}.

\section{Discussion}
\label{s:disc}

\subsection{Environmental dependence of the stellar mass function}

We investigate the environmental dependence of the SMF, by examining passive and star-forming galaxies separately. By defining the environment in terms of the local number density of galaxies we find no dependence of the shape of the SMF of SF galaxies on the environment, confirming the results we found at higher redshift in the galaxy cluster M1206. 
In the field, the values of the parameters of the SF galaxies SMF are in the ranges $\mathrm{10.46\, <\, log(\mstar)\,<\,  10.96}$ and $\mathrm{-1.51\,< \alpha <\, -1.20}$ \citep{ilbert2010,muzzin2013b}, rather similar to those we found in the external cluster region ($\mathrm{r\, >\,} \rtwo$). However, this comparison is not straightforward. In fact, the field SMF of SF galaxies changes with redshift and A209 is close to the lowest limit of the considered redshift intervals in \citet{ilbert2010}, and \citet{muzzin2013b}.\\
On the other hand, the SMF of passive galaxies depends on the local density, with a higher $\mstar$ and a flatter slope in the densest region, as shown in Fig.~10. The higher values of  $\mstar$ found in the central region of the cluster could be the result of the dynamical friction processes are effective in the cluster center, as also confirmed by the analysis of the stellar mass density profile (see below). Moreover, the SMF of passive galaxies is better fitted by a double Schechter function. This suggest the existence of two populations of passive galaxies: one formed by galaxies of high mass, while the other formed by galaxies of low masses ($\mste \mathrm{\, \leq 10^{9.5}\msun}$). In agreement with \cite{peng2010}, we can interpret the SMF as the combination of the SMF of mass quenched galaxies and the SMF of SF galaxies turned into passive by "environmental quenching". \\
The "environmental quenching" responsible of the double component of the SMF can be the effect of different mechanisms, like starvation,  ram-pressure stripping, galaxy-galaxy collisions and mergers, and harassment. 
Starvation and ram-pressure stripping affect the gas content of cluster galaxies, leading to a quenching of star formation. Starvation refers to the stripping of the external gas reservoir of the galaxy \citep{larson1980}, while ram-pressure strips the galaxy disk gas due to interaction with the hot and dense intra-cluster medium \citep{gunngott1972}. Starvation operates throughout the cluster, while ram-pressure stripping is strictly related to the ICM density distribution, so it is mostly effective in the central cluster region but can operate also out to the cluster virial radius \citep{tonnesen2007, tonnesen2014}. 
 Galaxy-galaxy collisions and mergers are more likely to be effective in the external cluster regions, where the relative speed of the encounters is not very high (\citealt{spitzer1951}, \citealt{negroponte1983}). However, the velocity dispersion of A209 remains very high also beyond the virial radius ($\mathrm{> 700\, km s^{-1}}$, Sartoris et al in prep.).\\
Harassment transforms galaxies by multiple high-speed galaxy-galaxy encounters and the interaction with the potential of the cluster as a whole (\citealt{moore1996, moore1998, moore1999}). \cite{moore1998} showed that harassment has no effect on systems as dense as giant ellipticals or spiral bulges but disk galaxies can be turned into spheroidal. This could induce a loss  of Sc or Sd-Im in the LF (at $\mathrm{\sim \, 2 \, mag\, below\, }L^*$, \citealt{binggeli1988}) and a comparable break in the spheroidal LF that lies $\mathrm{\sim \, 4 \, mag\, below\,  }L^*$. This would correspond to a dip in the SMF around $\mste \mathrm{\, =\, 10^{10.0} \, \msun }$ and an upturn at $\mste \mathrm{\, \leq\, 10^{9.0} \, \msun}$, in agreement with our findings in the less dense region of the cluster, suggesting that harassment could be the main driver of galaxy transformations in the external cluster regions. Recent simulations (\citealt{smith2010,bialas2015,smith2015}) indicate that harassment is not very effective  for newly infalling galaxies  with an orbital apocenter beyond half a virial radius. However, the low-mass passive galaxies in A209 are likely to have completed more than a single orbit.
In agreement with our results, \cite{popesso06} show that the $r$-band LF of local early-type galaxies requires a double Schechter function, with the faint-end slope having a significant and continuous variation with the environment, flattening in the central region of the cluster. They also interpret these results as an effect of galaxy transformation from star forming to quiescent systems through harassment in the periphery and dwarf tidal disruption in the core. On the other hand, not for all clusters there is significant evidence of a double Schechter LF (e.g. \citealt{rines2008}).   \\
Since galaxies experience different cluster environments when moving along their orbits, it is likely that their properties are not affected by just one  environmental process. For example, starvation can be effective in quenching the star formation of cluster galaxies even at large distances from the center, and part of the structural transformation could happen by preprocessing in groups before galaxies enter the cluster environment \citep{balogh2000}.\\
The evidence for two components in the SMF of red passive galaxies was first discovered in the field by \cite{drory2009} and has been recently shown at low (\citealt{baldry2012}) and high redshift (\citealt{muzzin2013b, ilbert2013, mortlock2015}). The mass at which the composition of the two populations becomes evident is higher at lower redshifts. At $z\mathrm{\, = \, 1}$ the signature of the double Schechter (an upturn of the SMF) is found at $\mste \mathrm{\, \sim \, 10^{9.0} \, \msun}$, while at $z\mathrm{\, = \, 0}$ this transition mass is higher, $\mste \mathrm{\, \sim \, 10^{10.0} \, \msun}$. According to this scenario, in fact, we did not find any evidence of a double Schechter in the SMF of passive galaxies in the cluster of galaxies M1206 at \textit{z}=0.44. In fact, at this redshift the transition mass is lower than the mass limit we had for that sample ($\mste \mathrm{\, \sim \, 10^{9.5} \, \msun}$). 
On the other hand, in the most dense region of the cluster, the second component of the SMF of passive galaxies seems to be suppressed, due to a lack of subgiant galaxies. This could be interpreted as the effect of processes responsible for the stripping of both the gas around the galaxies and their stellar content. These processes are probably tidal interactions with the cluster potential.

\subsection{The origin of the intra-cluster light}
By comparing the  single Schechter parametrization of the SMF in the central and most dense region of the cluster, with the ones obtained in outer regions, we find that the shape of the SMF of passive galaxies depends on the environment only in the most dense region of the cluster (Region (a)). \cite{annunziatella2014} found that there was a significant change in the slope SMF of passive galaxies in the cluster M1206 in the region corresponding to $\mathrm{R \leq 0.25\,} \rtwo$. This environmental effect observed in M1206 was interpreted in terms of formation of the ICL. In fact, as supported by the analysis of the ICL in that cluster (\citealt{Presotto+14}), \cite{annunziatella2014} discussed that the change in slope in the most dense regions of the cluster was due to tidal disruption of subgiant galaxies ($\mste \mathrm{\, \leq \,  10^{10.5} \msun}$) leading to the formation of at least part of the ICL. This scenario is supported also in A209 by the evaluation of the missing mass from subgiant galaxies ($\mathrm{\Delta M_{sub}}$) by the comparison with the ICL mass (Sect.~\ref{ss:icl_p}). According to the evaluation of the integral in Eq.~\ref{e:msub}, one of the main channel of formation of the ICL is the disruption of passive galaxies in the mass range $\mathrm{10^{9.0}\, < \, } \mste \mathrm{\, < 10^{10.0}\,  \msun}$, consistently with the analysis of the color of the ICL. 

\subsection{The differential evolution of high and low-mass galaxies}
We studied the mass-size relation of passive ETGs, separating the sample in low ($\mste\mathrm{ \, \leq\, 10^{10}\, \msun}$) and high-mass galaxies ($\mste \mathrm{\, \ge\, 10^{10}\, \msun}$). In fig.~18 we showed that low-mass galaxies follow a flatter relation respect to the one of high-mass galaxies. The relation of low-mass early-type passive galaxies is in agreement, within 1$\mathrm{\sigma}$, with the relation presented for star-forming galaxies at similar redshift in \cite{vanderWel2014}. Moreover, \cite{vanderWel2014_2} showed that at low redshifts emission-line galaxies are predominantly oblate and flat, that is, they are disks. This suggest that the mechanism responsible for the quenching of star-forming galaxies should also affect the galaxy morphology changing late types into early ones. Furthermore, since we found that the slope of the passive SMF becomes steeper in low-density regions, i.e. the number of low-mass passive galaxies increases in the external regions of the cluster, the quenching mechanism has to be effective also at large radii. 
This, together with the possible change of the galaxy morphology, could favor mechanisms such as the harassment,  (e.g. \citealt{treu2003}). However, we can not exclude that starvation of the gas reservoir could lead to a changing appearance of the galaxy, by reducing the prominence of the disk relative to the bulge.  
Other quenching mechanisms, which act mainly on the truncation of star-formation, without affecting morphologies, could be effective also at small radii. In fact, we find a small number of galaxies in the spectroscopic sample which are passive according to the spectral classification but have a late-type morphology ($n\mathrm{\, < \,2.5}$). The fraction of this type of galaxies is around 10\% in the cluster center, than decreases at larger radii.\\
The evolution of galaxies in different cluster regions proposed so far is also supported by the analysis of the orbits, according to which, low-mass galaxies ($\mste \mathrm{\, \leq \, 10^{10} \, \msun}$) in the center of the cluster have tangential orbits, meaning that they avoid small pericenters around the BCG. This could be explained by assuming that low-mass galaxies with radial orbits that go too close to the BCG are destroyed by tidal interactions. This is consistent with the analysis of the SMF and of the ICL discussed above. We also find that low-mass galaxies within $\mathrm{\sim \, }\rmtwo$ (0.65 Mpc) have smaller sizes per given mass than low-mass galaxies outside $\rmtwo$  (Fig.~\ref{f:res_ms}). We can suppose that low-mass galaxies in the center are influenced by tidal interactions that reduce their sizes.

\subsection{Evidence of dynamical friction in stellar mass density profile}
The environmental dependence of the stellar mass function is consistent with the relation between the number density, the stellar mass density and the total mass density profiles. We find that the ratio between the stellar mass density and the number density profiles is peaked towards the cluster center (fig.~20). This means that in the center of the cluster the mean mass is higher that in outer regions, a mass segregation effect. This effect is probably the combination of lack of low-mass galaxies and an excess of high-mass galaxies at small radii. The analysis of the ratio between the stellar mass density and the total mass density profiles suggests that dynamical friction has been effective in the cluster center, moving the most massive galaxies to the cluster center. Dynamical friction could also help lowering the galaxy velocities, thereby favoring the occurrence of mergers.\\

\section{Conclusions}
\label{s:conc}
In this paper, we have studied the SMF of member galaxies in the \textit{z}=0.209 cluster A209 from the CLASH-VLT ESO large programme. Our analysis has been based on a sample of $\mathrm{\sim 2000}$ cluster members, 53\% of which are spectroscopically confirmed members, and the other 47\% are photometrically selected from Subaru/SuprimeCam BVRIz imaging. We have obtained the stellar masses of the cluster members through SED fitting with \texttt{MAGPHYS} (\citealt{dacunha2008}). With this same dataset, we have evaluated the mass density and the number density profiles of the cluster members, reaching a completeness mass limit of $\mathrm{\mste=10^{8.6} \msun}$. We also have studied the intra-cluster light distribution in this cluster, using  GALAPAGOS (\citealt{barden2012}) and GALtoICL (\citealt{Presotto+14}). We have compared our results with those reported in \cite{annunziatella2014} and \cite{Presotto+14} relative to the cluster M1206 at \textit{z}=0.44.

The results of our study can be summarized as follows:

\begin{itemize}

\item The SMFs of star-forming and passive member galaxies are significantly different in all cluster regions. The SMF of star-forming galaxies does not show any  environmental dependence.
At variance with the SMF of SF galaxies, the SMF of passive galaxies shows a double component, that becomes more evident in lower density regions of the cluster. We suggest that this double component originates from the environmental quenching of SF galaxies. This quenching becomes more
effective at lower masses with increasing cosmic time, as indicated by the difference between the
SMFs of passive galaxies in A209 and in the $\textit{z}=0.44$ cluster M1206. 
\item Additional support for environmental quenching of SF galaxies is provided by the flattening of the mass-size relation for ETGs at low masses ($\mste \mathrm{\leq  10^{10}\, \msun}$). At high masses, the mass-size relation slope is indistinguishable from that for field passive ETGs, while at low masses, the slope is similar to that for field SF galaxies. 
We interpret this as evidence for quenching and morphological transformation of SF galaxies
accreted from the field. From the analysis of the mass-size relation we also find
that low-mass ETGs become more compact in the central cluster region ($r\, \leq\, \rmtwo$), a possible signature of the effect of tidal interactions.
\item We find evidence of environmental dependence of the SMF of passive galaxies. In particular, the SMF of passive galaxies in the highest density region is  characterized by a flatter slope at the low-mass end than the SMFs in lower density regions. This indicates a lack of low-mass galaxies, that could be the consequence of tidal stripping/destruction in the densest cluster region. This process could be at the origin of the ICL. The mass of the ICL is in fact comparable to that of the low-mass galaxies that are missing in the densest cluster region. 
The main contributors to the ICL mass ($\mathrm{\sim 50 \%}$ of the total) are galaxies in the mass range $\mathrm{10^{9.0}\, < \,} \mste \mathrm{\, < 10^{10.0}\,  \msun}$.
Moreover, the ICL color is similar to that of passive spectroscopic members. 
\item Additional support for tidal destruction of low-mass galaxies comes from the analysis of their orbits in the cluster. In fact, low-mass galaxies are characterized by tangential orbits that avoid small pericenters, presumably because those that cross the cluster center get tidally destroyed, whereas more massive galaxies ($\mste \mathrm{\, > \,10^{10} \msun}$)  show radial orbits thus suggesting that they can penetrate through dense cluster inner environments. 
\item The SMF of passive galaxies in the highest density region is also characterized by a higher $\mstar$ than the SMFs in lower density regions. This indicates an excess of massive galaxies, that may originate from dynamical friction. This interpretation is further supported by the comparative analysis of the cluster stellar and total mass density profiles. 
\end{itemize}
Finding a consistent scenario for cluster assembly is complex, as due to the interplay of a number of different physical processes effective in clusters. It requires a knowledge of the different contributions of the ICL and subgiant galaxies to the stellar mass budget, in combination with stellar population and structural properties of galaxies, as well as their kinematics.

In the future, we plan to extend this analysis to clusters at higher redshifts, and to carry out a detailed comparison of our results with predictions from semi-analytical models.

\begin{acknowledgements}
We thank the referee for her/his useful comments, which have helped us
to improve our paper.
We thank G. Rudnick for useful discussion.
We thank the ESO User Support group for the excellent support of the Large Programme 186.A-0798.
Based [in part] on data collected at Subaru Telescope and obtained from the SMOKA, which is
operated by the Astronomy Data Center, National Astronomical Observatory of Japan.
We acknowledge support from PRIN-INAF 2014 1.05.01.94.02 (PI M. Nonino).
PR acknowledges the hospitality and support of the visitor program of the DFG cluster of excellence "Origin and Structure of the Universe". 
R.D. gratefully acknowledges the support provided by the BASAL Center for Astrophysics and Associated Technologies (CATA), and by FONDECYT grant N. 1130528

\end{acknowledgements}

\bibliographystyle{aa}

\bibliography{bibliography}

\end{document}